\documentclass[11pt]{article}
\usepackage[margin=1in]{geometry}
\usepackage{amsmath,amssymb,bm,graphicx}
\usepackage{xcolor}
\usepackage{hyperref}
\usepackage{booktabs}
\definecolor{linkblue}{RGB}{24,78,119}
\hypersetup{
  colorlinks=true,
  linkcolor=linkblue,
  citecolor=linkblue,
  urlcolor=linkblue,
  pdfauthor={Anish Ghoshal},
  pdftitle={Thermal fluctuations at finite chemical potential in the early universe}
}
\allowdisplaybreaks[2]
\newcommand{\mpl}{M_{\rm Pl}}
\newcommand{\cH}{\mathcal H}
\newcommand{\Pz}{\mathcal P_{\zeta}}
\newcommand{\PS}{\mathcal P_{S}}
\newcommand{\dd}{\mathrm d}
\newcommand{\order}{\mathcal O}
\title{Statistical Fluctuations as Primordial Correlators in the CMB:\\ \it{Finite Chemical Potential Thermodynamics}}
\date{}

\author{Anish Ghoshal$^{1}$, Anupam Mazumdar$^{2}$, Bart\l omiej Sikorski$^{3}$ \\[0.6em]  {\small $^{1}$Department of Astronomy and Physics, University of Sussex, Brighton BN1 9QH, United Kingdom }  \\ {\small $^{2}$ Canadian Institute of Theoretical Astrophysics, University of Toronto, Toronto, Canada} \\{\small $^{3}$Faculty of Physics, University of Warsaw, Pasteura 5, 02-093 Warsaw, Poland }\\[0.6em] {\small \href{mailto:a.ghoshal@sussex.ac.uk} {\texttt{a.ghoshal@sussex.ac.uk}} \quad  \quad \href{mailto:bartlomiej.sikorski@fuw.edu.pl} {\texttt{bartlomiej.sikorski@fuw.edu.pl}} } } 
\begin{document}
\maketitle

\begin{abstract}
Can primordial cosmological correlations originate from the unavoidable statistical fluctuations of an early-universe thermal system? We develop a systematic framework that connects grand-canonical thermodynamics, stochastic transport, and gauge-invariant cosmological perturbations for a charged fluid at finite chemical potential. The full energy-charge susceptibility matrix determines the primordial curvature, charge isocurvature, and cross-correlation amplitudes, while transport specifies when each mode ceases to track equilibrium. For an isolated, adiabatic, extensive conformal plasma with conserved charge-to-entropy ratio and power-law diffusion, we derive a robust and transport-independent result: diffusive freeze-out produces the universal blue spectrum $\mathcal P_S(k)\propto k^3$, with $n_{\rm iso}=4$, and the equal-time local-equilibrium curvature-isocurvature covariance vanishes in the conformal source basis. This establishes a sharp obstruction: neither finite chemical potential nor replacing Hubble crossing by diffusion crossing is sufficient by itself to generate an approximately scale-invariant spectrum. We then identify a concrete route around this obstruction by considering an open subsystem of effectively massless Dirac fermions during a sourced quasi-de Sitter phase. Along this non-adiabatic trajectory, grand-canonical energy cumulants generate a nearly scale-invariant red curvature spectrum. A representative benchmark reproduces $\mathcal P_{\zeta_X}(k_\star)=2.10\times10^{-9}$ and $n_s(k_\star)=0.9649$, remains within approximately $0.30\%$ of the fitted power law over $0.002\leq k/k_\star\leq4$, and predicts weak negative running together with a small positive intrinsic non-Gaussian amplitude, from higher-order cumulants, $f_{\rm NL}(k_\star)\simeq0.106$ and $g_{\rm NL}(k_\star)\simeq0.0158$. With an independently imposed two-helicity vacuum tensor spectrum, the corrected    tensor-to-scalar ratio is $r_{t/s}(k_\star)=1.82\times10^{-4}$. The comparison isolates the limitation of isolated conformal thermal seeding and the open-system ingredient needed to overcome it. Our results make thermal energy-charge fluctuations a calculable source of realistic primordial scalar correlations, while identifying the reservoir dynamics, equilibration, and later curvature-isocurvature transfer that a microscopic model must supply.
\end{abstract}

\section{Introduction}
\label{sec:intro}

The observed cosmic microwave background is remarkably close to a Gaussian, adiabatic random field with an almost scale-invariant and slightly red scalar spectrum. In the standard account, these correlations originate from quantum vacuum fluctuations stretched beyond the Hubble radius during accelerated expansion. Yet the early universe was also a many-body system, and every finite thermal system carries irreducible statistical fluctuations. This raises a basic question: can equilibrium fluctuations of energy and conserved charge provide a quantitatively controlled route to primordial curvature and isocurvature perturbations? The question is especially timely because thermal mechanisms tie the statistics of primordial fluctuations directly to microphysical quantities such as susceptibilities, transport coefficients, chemical potentials, relaxation rates, and interaction-induced correlations.

A useful starting point is the familiar canonical relation. For a neutral canonical system, the relation $\langle(\delta E)^2\rangle=T^2C_V$ relates the energy variance to the heat capacity. Ref.~\cite{Biswas:2013lna} developed a general prescription for converting statistical thermal fluctuations in a single fluid into scalar and tensor spectra and higher cumulants. Before that, many authors contributed towards our understanding of density fluctuations in the cosmic microwave background radiation through thermal fluctuations, see~\cite{Cai:2009hc,Magueijo:2002pg, Nayeri:2005ck,Biswas:2007hagedorn, Magueijo:2007loop}.

Their analysis also exposes the central obstacle: an extensive, adiabatic, dominant thermal fluid with constant equation of state generically produces a strongly blue dimensionless spectrum, with $n_s=4$ in the simplest limit. Thermal alternatives must therefore explain not only how fluctuations are generated, but also how equilibrium scaling, freeze-out, and gravitational conversion combine to avoid this blue result.

Previous thermal scenarios evade this scaling in several different ways. In bouncing cosmologies, the background equation of state and the passage through a nonsingular bounce change the relation between thermal length scales and late-time curvature perturbations \cite{Cai:2009hc}. String-gas cosmology exploits Hagedorn thermodynamics and holographic scaling rather than an ordinary point-particle extensive gas \cite{Brandenberger:2011et}. Thermal or cyclic inflation can convert temperature fluctuations at the end of an accelerated phase, and phase transitions can amplify higher thermodynamic derivatives and non-Gaussianity \cite{LythStewart:1996thermal,Biswas:2013phase,Bae:2025thermal}. These constructions demonstrate that thermal seeding is possible in principle, but they rely on thermodynamics, backgrounds, or conversion surfaces that differ substantially from an isolated relativistic plasma.

Warm inflation provides a closer dynamical precedent. Dissipative interactions can maintain a thermal bath during accelerated expansion, and stochastic thermal fluctuations can then contribute directly to the scalar spectrum \cite{Berera:1995warm,Hall:2004warm,BereraMossRamos:2009review}. Modern microscopic realizations show that light fermions and chemical responses can materially affect both dissipation and noise \cite{BasteroGil:2016warm,Berghaus:2025SM,Broadberry:2026chemical}. The present construction is related to this literature through its continuously sourced thermal component, but it asks a more specific thermodynamic question. Rather than beginning with inflaton noise, it begins with the grand-canonical covariance of energy and charge in a dark subsystem and tracks how that covariance is frozen and converted into cosmological perturbations.

Finite chemical potential changes the problem qualitatively. The equilibrium state is no longer characterized by a single energy variance: it contains an energy-charge susceptibility matrix with a generally nonzero mixed covariance. A local fluctuation in charge can therefore carry energy, and the same thermal state can seed adiabatic, charge isocurvature, and correlated modes. This places the problem naturally within multifluid cosmological perturbation theory, where gauge-invariant entropy modes describe displacements transverse to the homogeneous trajectory \cite{Bucher:2000general,Malik:2008im}. It also makes transport indispensable. Conserved charge relaxes diffusively, so each comoving mode ceases to track equilibrium when its physical diffusion rate becomes comparable to the expansion rate. Relativistic fluctuating hydrodynamics fixes the corresponding noise through fluctuation-dissipation and clarifies when causal or non-Markovian corrections are required \cite{Kovtun:2012rj,Kapusta:2012hydro,Crossley:2017eft,Basar:2024qog}.

For clarity, we keep three parts of the calculation separate. First, grand-canonical thermodynamics fixes the equal-time covariance and higher connected cumulants of energy and charge. Second, stochastic transport fixes the mode-dependent freeze-out scale and determines which part of the equilibrium covariance survives. Third, gauge-invariant gravitational evolution projects the frozen variables onto curvature and isocurvature perturbations and allows later entropy-to-curvature conversion. This separation is important because a realistic amplitude or tilt cannot be inferred from thermodynamics alone: the result depends equally on the background trajectory, relaxation dynamics, and transfer history.

Two complementary regimes are developed. The first is an isolated conformal plasma with a conserved charge, constant charge-to-entropy ratio, and power-law diffusion. In this limit we obtain a sharp result: the dimensionless charge isocurvature spectrum obeys $\mathcal P_S\propto k^3$, independently of the diffusion exponent, and its equal-time cross-correlation with curvature vanishes. Thus finite chemical potential and diffusive crossing do not by themselves solve the blue-spectrum problem. The result identifies precisely what must change: conformality, extensivity, the conserved $\mu/T$ trajectory, power-law local transport, or the post-freeze-out transfer.

The second regime is an open dark subsystem of effectively massless Dirac fermions during a sourced quasi-de Sitter phase. The subsystem exchanges energy and charge with a reservoir, its physical chemical potential is approximately constant over a finite interval, and $\mu/T$ therefore evolves. These assumptions deliberately violate those behind the conformal no-go result. For a representative benchmark, the construction reproduces the observed scalar amplitude and red tilt at the pivot and remains close to a power law across the fitted range. The same thermodynamic cumulants predict weak running and small higher-order amplitudes. The two regimes play different roles. The isolated sector gives a clean obstruction, whereas the open sector shows how it can be avoided and exposes the energy and charge sources required to do so.

The paper is organized as follows. Section~\ref{sec:thermo} develops the grand-canonical susceptibility formalism. Section~\ref{sec:pert} constructs gauge-invariant curvature and charge-entropy modes. Sections~\ref{sec:diff} and \ref{sec:spectra} formulate stochastic diffusion and the freeze-out covariance. Section~\ref{sec:conformal} derives the universal blue spectrum in the conformal conserved-charge limit. Sections~\ref{sec:open_curvature} and \ref{sec:dark_fermion} develop the open-subsystem curvature channel and the dark-fermion realization. Section~\ref{sec:conversion} discusses subsequent evolution and phenomenology, followed by the synthesis in Sec.~\ref{sec:discussion}.

Throughout, natural units $\hbar=c=k_{\rm B}=1$ are used. The symbol $T$ denotes the local matter temperature, $\mu_a$ the chemical potential associated with charge $N_a$, $a(t)$ the scale factor, $H=\dot a/a$ the physical Hubble rate, and $\mathcal H=aH$ the conformal Hubble rate. An overdot denotes differentiation with respect to proper time $t$, while a prime denotes differentiation with respect to conformal time $\eta$. The reduced Planck mass is written as $\mpl=(8\pi G)^{-1/2}$ unless the unreduced mass $M_{\rm Pl}$ is displayed explicitly. Bold symbols denote spatial vectors or vectors in thermodynamic state space; the intended meaning is stated where each appears. Connected expectation values carry the subscript $c$.
 {For reference, $\rho$ and $p$ denote total energy density and pressure unless an explicit component label such as $X$ is attached; $n$ is a physical charge-number density; $s$ is the entropy density; $T$ and $\mu$ are the local temperature and chemical potential; $k$ is a comoving wavenumber and $k_{\rm ph}=k/a$ its physical value.  The symbols $\Sigma_{AB}$ and $\mathcal C_{IJ}$ denote, respectively, a dimensional thermodynamic covariance density and the covariance of dimensionless cosmological variables.  Conductivity, diffusion coefficient, microscopic correlation length, and relaxation time are denoted by $\sigma$, $D$, $\xi$, and $\tau$, respectively.}

\section{Grand-canonical thermodynamics and fluctuation covariances}
\label{sec:thermo}

We begin with the equilibrium quantities needed in the stochastic and cosmological analysis. The essential object is the covariance density $\Sigma_{AB}$, which measures fluctuations per unit physical volume. It should be distinguished from the smoothed covariance $\mathcal C_{IJ}$ of the dimensionless curvature and entropy variables introduced in Sec.~\ref{sec:pert}.

\subsection{Generating functional and covariance matrix}

Consider a homogeneous fluid in a physical volume $V$ with Hamiltonian $E$ and conserved charges $N_a$. Its grand-canonical partition function is \cite{LandauLifshitz:1980statphys}
\begin{equation}
 Z(\beta,\alpha_a)=\mathrm{Tr}\exp\left[-\beta E+\sum_a\alpha_a N_a\right],
 \qquad \alpha_a\equiv\beta\mu_a,
\end{equation}
where $\beta=T^{-1}$.

The intensive variables $\alpha_a=\mu_a/T$ are the natural sources for the conserved charges. Differentiating at fixed $\alpha_a$, rather than at fixed $\mu_a$, is necessary because the grand-canonical Boltzmann weight is $\exp[-\beta E+\alpha_aN_a]$. The composite index $A$ runs over energy and all conserved charges, so repeated thermodynamic indices label matrix components rather than spacetime directions.
\begin{equation}
 \lambda_A=(-\beta,\alpha_a),
 \qquad Q_A=(E,N_a).
\end{equation}
The Massieu function $\Psi=\ln Z=\beta pV$ generates the mean densities $q_A=(\rho,n_a)$ and their connected cumulants.  This source-derivative construction is the equilibrium limit of relativistic fluctuating hydrodynamics and fixes the static covariance that the Langevin theory must reproduce through fluctuation-dissipation \cite{Kovtun:2012rj,Kapusta:2012hydro}:
\begin{align}
 \label{eq:SigmaDef}
 q_A&=\frac{1}{V}\frac{\partial\Psi}{\partial\lambda_A},\\
 \Sigma_{AB}&\equiv \frac{1}{V}\langle\delta Q_A\delta Q_B\rangle_c
 =\frac{1}{V}\frac{\partial^2\Psi}{\partial\lambda_A\partial\lambda_B}.
\end{align}
Thermodynamic stability requires $\Sigma$ to be positive semidefinite. For one conserved charge,
\begin{equation}
 \Sigma=
 \begin{pmatrix}
 \Sigma_{\rho\rho}&\Sigma_{\rho n}\\
 \Sigma_{\rho n}&\Sigma_{nn}
 \end{pmatrix},
 \qquad
 \det\Sigma\ge0.
\end{equation}
The off-diagonal entry has a direct physical interpretation: a charge fluctuation generally carries energy and therefore correlates the adiabatic and isocurvature modes.

Here $\Sigma_{\rho\rho}$ is the energy-density variance density, $\Sigma_{nn}$ is the charge-density variance density, and $\Sigma_{\rho n}$ is their covariance density. Positive semidefiniteness means that every real linear combination of energy and charge has nonnegative variance. The determinant condition therefore supplies both a stability test and a useful check on numerical equations of state.

For a window $W_R(\bm x)$ normalized by $\int\dd^3x\,W_R=1$, define $\delta q_{A,R}=\int\dd^3x\,W_R\delta q_A$. If $R$ is much larger than the microscopic correlation length $\xi$, extensivity gives  \cite{LandauLifshitz:1980statphys,Forster:1990hydro}
\begin{equation}
 \langle\delta q_{A,R}\delta q_{B,R}\rangle
 =\frac{\Sigma_{AB}}{V_R^{\rm eff}},
 \qquad
 \frac{1}{V_R^{\rm eff}}\equiv\int\dd^3x\,W_R^2(\bm x).
 \label{eq:windowvariance}
\end{equation}
 {Equation~\eqref{eq:windowvariance} follows by inserting the local-equilibrium correlator $\langle\delta q_A(\bm x)\delta q_B(\bm y)\rangle=\Sigma_{AB}\delta^{(3)}(\bm x-\bm y)$ into the definition of the smoothed variables.  The two window integrals collapse to $\int W_R^2$, which defines $V_R^{\rm eff}$.  Thus the variance decreases as the inverse number of independent correlation cells in the averaging volume, as expected for an extensive system with $R\gg\xi$ \cite{LandauLifshitz:1980statphys,Forster:1990hydro}.}
Equivalently, the long-wavelength physical Fourier spectrum is white,

Here ``white'' refers to the dimensional spectrum, which is independent of $k_{\rm ph}$ at leading order. The dimensionless spectrum is nevertheless proportional to $k_{\rm ph}^3$. The correction $\order(k_{\rm ph}^2\xi^2)$ records the leading failure of locality when the physical wavelength approaches the microscopic correlation length $\xi$.
\begin{equation}
 \langle\delta q_A(\bm k_{\rm ph})\delta q_B(\bm k'_{\rm ph})\rangle
 =(2\pi)^3\delta^{(3)}(\bm k_{\rm ph}+\bm k'_{\rm ph})
 \left[\Sigma_{AB}+\order(k_{\rm ph}^2\xi^2)\right].
 \label{eq:white}
\end{equation}
 {Here ``white'' means that the dimensional spectrum approaches the constant matrix $\Sigma_{AB}$ as $k_{\rm ph}\xi\to0$.  Locality makes the first analytic correction quadratic in $k_{\rm ph}\xi$ for an isotropic, parity-even medium.  Multiplication by the phase-space factor $k_{\rm ph}^3/(2\pi^2)$ then makes the corresponding dimensionless spectrum blue, even though the dimensional spectrum is flat \cite{Forster:1990hydro,Kovtun:2012rj}.}

\subsection{Explicit susceptibilities and changes of variables}

The entries of $\Sigma$ can be written explicitly in terms of standard thermodynamic derivatives.  Holding $\alpha=\mu/T$ fixed when differentiating with respect to $\beta$ is essential.  From $\Psi=\beta pV$ one obtains
\begin{align}
 \Sigma_{nn}&=T\left(\frac{\partial n}{\partial\mu}\right)_T
 \equiv T\chi_{TT},\label{eq:Snnexplicit}\\
 \Sigma_{\rho n}&=T^2\left(\frac{\partial n}{\partial T}\right)_{\alpha}
 =T^2\left[\left(\frac{\partial n}{\partial T}\right)_\mu
 +\frac{\mu}{T}\left(\frac{\partial n}{\partial\mu}\right)_T\right],\label{eq:Srhonexplicit}\\
 \Sigma_{\rho\rho}&=T^2\left(\frac{\partial\rho}{\partial T}\right)_{\alpha}
 =T^2\left[\left(\frac{\partial\rho}{\partial T}\right)_\mu
 +\frac{\mu}{T}\left(\frac{\partial\rho}{\partial\mu}\right)_T\right].\label{eq:SrrExplicit}
\end{align}
 {To obtain Eqs.~\eqref{eq:Snnexplicit}-\eqref{eq:SrrExplicit}, one differentiates $\Psi(\beta,\alpha)=\beta pV$ in its natural sources.  Since $\partial_{(-\beta)}=T^2\partial_T|_\alpha$ and $\partial_\alpha=T\partial_\mu|_T$, the source Hessian directly generates the charge variance, mixed energy-charge covariance, and energy variance.  The chain rule $\partial_T|_\alpha=\partial_T|_\mu+(\mu/T)\partial_\mu|_T$ gives the second forms.  These standard grand-canonical fluctuation identities are reviewed in Refs.~\cite{LandauLifshitz:1980statphys,Laine:2016thermal}.}
Here $\chi_{TT}\equiv(\partial n/\partial\mu)_T$ is the static charge susceptibility.

 Thus $(\partial n/\partial T)_\alpha$ follows the ray of constant $\mu/T$, while $(\partial n/\partial T)_\mu$ follows a trajectory of constant physical chemical potential. This distinction becomes central when comparing the isolated conformal sector with the sourced open subsystem.  The notation does not mean a temperature-temperature correlator; the subscript only reminds us that $T$ is held fixed.  Equations~\eqref{eq:Snnexplicit}-\eqref{eq:SrrExplicit} follow directly by differentiating $\Psi$ in the natural variables $(\beta,\alpha)$.  They also make the dimensions transparent: in $3+1$ dimensions $[\Sigma_{nn}]=E^3$, $[\Sigma_{\rho n}]=E^4$, and $[\Sigma_{\rho\rho}]=E^5$.

Numerical applications often begin with a tabulated equation of state expressed in the variables $(T,\mu)$.  Define the Hessian of pressure
\begin{equation}
 H_p=\begin{pmatrix}p_{,TT}&p_{,T\mu}\\p_{,\mu T}&p_{,\mu\mu}\end{pmatrix},
 \qquad p_{,\mu\mu}=\chi_{TT}.
\end{equation}
 {In Eq.~(11), $H_p$ is the Hessian matrix of the pressure with respect to $(T,\mu)$.  Its entries measure the linear response of entropy and charge densities to changes of temperature and chemical potential: $p_{,TT}=\partial s/\partial T|_\mu$, $p_{,T\mu}=\partial n/\partial T|_\mu$, and $p_{,\mu\mu}=\partial n/\partial\mu|_T$.  Equality of the mixed derivatives is the Maxwell relation $\partial s/\partial\mu|_T=\partial n/\partial T|_\mu$ \cite{LandauLifshitz:1980statphys}.}
The thermodynamic stability requirements are positive heat capacity at fixed charge and positive charge susceptibility.  In the grand-canonical representation they are equivalently encoded by
\begin{equation}
 \Sigma_{nn}\geq0,\qquad \Sigma_{\rho\rho}\geq0,
 \qquad \Sigma_{\rho\rho}\Sigma_{nn}-\Sigma_{\rho n}^2\geq0.
 \label{eq:stabilityineq}
\end{equation}
The last inequality is the Cauchy-Schwarz bound on energy-charge correlations.  Saturation means that only one independent thermodynamic fluctuation survives; in that limit one linear combination of curvature and isocurvature has zero equilibrium variance.

A second useful basis consists of the entropy density $s$ and the charge yield $Y=n/s$.

The yield $Y$ is dimensionless and measures charge per unit entropy. It remains constant after both the charge and the entropy in a comoving volume are conserved. This makes $\delta Y$ preferable to $\delta n/n$ near a charge-symmetric background, where the latter becomes singular only because its normalization tends to zero.  Linearizing,
\begin{equation}
 \delta Y=\frac{\delta n}{s}-\frac{n}{s^2}\delta s,
 \qquad
 \delta s=\frac{\delta\rho-\mu\delta n}{T},
 \label{eq:yieldfluct}
\end{equation}
 {The first relation in Eq.~\eqref{eq:yieldfluct} is the linear variation of the ratio $Y=n/s$.  The second follows from the local first law $\dd\rho=T\dd s+\mu\dd n$ at fixed physical volume.  Hence $\delta Y$ removes the part of a charge fluctuation caused solely by an adiabatic entropy fluctuation and remains finite when the background charge tends to zero \cite{LandauLifshitz:1980statphys}.}
where the second expression is the first law at fixed physical volume.  The yield basis remains regular when the charge contributes negligibly to the total energy density, and it is the natural basis after chemical decoupling because the homogeneous $Y$ is conserved in an adiabatically expanding universe.

\subsection{Thermodynamic representation}

The pressure determines
\begin{equation}
 s=\left(\frac{\partial p}{\partial T}\right)_{\mu_a},
 \qquad
 n_a=\left(\frac{\partial p}{\partial\mu_a}\right)_T,
 \qquad
 \rho=-p+Ts+\sum_a\mu_a n_a.
 \label{eq:thermoid}
\end{equation}
Although one may transform Eq.~\eqref{eq:SigmaDef} to the $(T,\mu_a)$ Hessian of $p$, the source basis $(-\beta,\beta\mu_a)$ is preferable because it gives the covariance matrix without ambiguous factors of $T$. Higher connected cumulants follow from additional source derivatives,
\begin{equation}
 \Sigma_{A_1\cdots A_m}
 =\frac{1}{V}\frac{\partial^m\Psi}
 {\partial\lambda_{A_1}\cdots\partial\lambda_{A_m}},
 \label{eq:highercumulants}
\end{equation}
providing a direct route to mixed adiabatic-isocurvature non-Gaussianity.  The interpretation of these derivatives as local hydrodynamic noise cumulants, and their nonlinear evolution in an expanding relativistic fluid, is discussed in Refs.~\cite{Kapusta:2012hydro,Basar:2024qog,An:2021wof}.

\subsection{Magnitude and interpretation of equilibrium fluctuations}

For a spherical top-hat region of radius $R$, $V_R=4\pi R^3/3$, the root-mean-square fractional charge fluctuation is
\begin{equation}
 \frac{\sqrt{\langle(\delta n_R)^2\rangle}}{|n|}
 =\left(\frac{T\chi_{TT}}{n^2V_R}\right)^{1/2}.
 \label{eq:fractionalchargeestimate}
\end{equation}
 {Equation~\eqref{eq:fractionalchargeestimate} combines $\Sigma_{nn}=T\chi_{TT}$ with the inverse-volume law in Eq.~\eqref{eq:windowvariance}.  It shows that the absolute fluctuation is controlled by the susceptibility, whereas division by a small mean density $n$ can make the fractional fluctuation large.  This is a normalization effect, not a thermodynamic singularity \cite{LandauLifshitz:1980statphys}.}
If a relativistic charge asymmetry is parametrized by $n=Y s$ with $s=(2\pi^2/45)g_{*s}T^3$, and $\chi_{TT}=c_\chi T^2$, then
\begin{equation}
 \frac{\sqrt{\langle(\delta n_R)^2\rangle}}{|n|}
 \simeq \frac{1}{|Y|g_{*s}}
 \left(\frac{45^2 c_\chi}{4\pi^4T^3V_R}\right)^{1/2}.
 \label{eq:asymmetryestimate}
\end{equation}
A very small homogeneous asymmetry therefore produces a large \emph{fractional} fluctuation even when the absolute fluctuation is perturbative.  This enhancement is not a divergence of the grand-canonical ensemble.  It instead indicates that $\delta n/n$ becomes an unsuitable variable as $n\to0$; the yield or the eventual relic energy density then provides a regular alternative.

\section{Gauge-invariant curvature and charge-entropy modes}
\label{sec:pert}

We next translate the local thermodynamic fluctuations into slicing-independent cosmological variables. The metric potentials $A$, $B$, $\psi$, and $E$ describe scalar perturbations of the lapse, shift, intrinsic spatial curvature, and scalar shear, respectively. No gauge choice is required for the final variables $\zeta$ and $S_a$.

We use the scalar-perturbation conventions of Refs.~\cite{Malik:2008im,Baumann:2009inflation} for a spatially flat FLRW metric,
\begin{equation}
 \dd s^2=a^2(\eta)\left[-(1+2A)\dd\eta^2
 +2\partial_iB\dd x^i\dd\eta
 +\left((1-2\psi)\delta_{ij}+2\partial_i\partial_jE\right)\dd x^i\dd x^j\right].
\end{equation}
The curvature perturbation on uniform total-density hypersurfaces is
\begin{equation}
 \zeta=-\psi-\cH\frac{\delta\rho}{\rho'},
 \label{eq:zeta}
\end{equation}
where a prime denotes a conformal-time derivative.

The variable $\zeta$ is the curvature perturbation evaluated on hypersurfaces of uniform total energy density. It is especially useful because it is conserved on super-Hubble scales for an adiabatic system with negligible anisotropic stress and negligible gradient terms. For a separately conserved charge current, $\nabla_\mu J_a^\mu=0$, the background number density obeys $n_a'+3\cH n_a=0$. We may therefore define
\begin{equation}
 \zeta_{n_a}=-\psi-\cH\frac{\delta n_a}{n_a'}
 =-\psi+\frac{\delta n_a}{3n_a}.
 \label{eq:zetan}
\end{equation}
 {Equation~\eqref{eq:zetan} is the curvature perturbation on hypersurfaces of uniform charge density.  The second equality uses separate background charge conservation, $n_a'=-3\mathcal H n_a$.  A patch with positive $\delta n_a$ therefore reaches a fixed-density hypersurface at a shifted local expansion, encoded by $\zeta_{n_a}$ \cite{Malik:2008im,Bucher:2000general}.}
The gauge-invariant charge isocurvature perturbation relative to the total energy density is
\begin{equation}
 S_a\equiv 3(\zeta_{n_a}-\zeta)
 =\frac{\delta n_a}{n_a}-\frac{\delta\rho}{\rho+p},
 \label{eq:Sdef}
\end{equation}
where the final equality uses $\rho'=-3\cH(\rho+p)$. Equation~\eqref{eq:Sdef} is valid on any slicing at linear order because the time-shift pieces cancel.  It is the conserved-charge analogue of the relative entropy modes used in the general classification of regular primordial adiabatic and isocurvature initial conditions \cite{Bucher:2000general,Malik:2008im}.

For one charge, define the fluctuation vector $\delta\bm q=(\delta\rho,\delta n)^T$ and projection vectors
\begin{equation}
 \bm u_S=\begin{pmatrix}-(\rho+p)^{-1}\\ n^{-1}\end{pmatrix},
 \qquad
 S=\bm u_S^T\delta\bm q.
 \label{eq:uS}
\end{equation}
On a spatially flat slicing, the curvature fluctuation is
\begin{equation}
 \zeta=\frac{\delta\rho}{3(\rho+p)}
 \equiv\bm u_\zeta^T\delta\bm q,
 \qquad
 \bm u_\zeta=\begin{pmatrix}[3(\rho+p)]^{-1}\\0\end{pmatrix}.
 \label{eq:uzeta}
\end{equation}
 {Equation~\eqref{eq:uzeta} uses the spatially flat slicing $\psi=0$ together with $\rho'=-3\mathcal H(\rho+p)$.  The covector $\bm u_\zeta$ simply selects the energy-density component of $\delta\bm q$ and divides it by the background enthalpy $3(\rho+p)$ \cite{Malik:2008im,Baumann:2009inflation}.}
The local-equilibrium covariance of $(\zeta,S)$ is consequently the projection of $\Sigma$,
\begin{equation}
 \mathcal C_{IJ}=\bm u_I^T\Sigma\bm u_J,
 \qquad I,J\in\{\zeta,S\}.
 \label{eq:projectedcov}
\end{equation}
 {Equation~\eqref{eq:projectedcov} is ordinary covariance propagation under a linear change of variables: if $X_I=u_I^A\delta q_A$, then $\langle X_I X_J\rangle=u_I^A\Sigma_{AB}u_J^B$.  No gravitational dynamics is added at this step; the equation only projects the thermodynamic fluctuation ellipse onto adiabatic and entropy directions.}
In particular,
\begin{align}
 \mathcal C_{SS}&=
 \frac{\Sigma_{\rho\rho}}{(\rho+p)^2}
 -\frac{2\Sigma_{\rho n}}{n(\rho+p)}
 +\frac{\Sigma_{nn}}{n^2},
 \label{eq:CSS}\\
 \mathcal C_{\zeta S}&=
 \frac{1}{3(\rho+p)}
 \left[-\frac{\Sigma_{\rho\rho}}{\rho+p}
 +\frac{\Sigma_{\rho n}}{n}\right].
 \label{eq:CzS}
\end{align}
 {Eqs.~\eqref{eq:CSS} and \eqref{eq:CzS} follow by explicitly multiplying the two-component vectors in Eqs.~\eqref{eq:uS} and \eqref{eq:uzeta}.  The three terms in $\mathcal C_{SS}$ are the energy contribution, the mixed interference term, and the charge contribution.  The sign of $\mathcal C_{\zeta S}$ is therefore fixed by whether a typical positive charge fluctuation carries more or less energy than the adiabatic ratio $n/(\rho+p)$.}
These equations provide model-independent thermodynamic predictions for both the amplitude and the sign of the primordial curvature-isocurvature correlation, subject to the dynamical freeze-out transfer derived below.

The vectors $\bm u_\zeta$ and $\bm u_S$ are projection covectors in the two-dimensional fluctuation space $(\delta\rho,\delta n)$. Equation~\eqref{eq:projectedcov} is therefore a change of basis: it contains no additional dynamics. All wavelength dependence enters later through the freeze-out temperature, diffusion coefficient, and transfer functions.

\subsection{Gauge transformations and conservation}

Under an infinitesimal scalar time shift $\eta\rightarrow\eta+\xi^0$, matter perturbations transform as
\begin{equation}
 \widetilde{\delta\rho}=\delta\rho-\rho'\xi^0,
 \qquad
 \widetilde{\delta n}=\delta n-n'\xi^0,
 \qquad
 \widetilde\psi=\psi+\cH\xi^0.
\end{equation}
 {These transformations are the Lie-dragging of background scalars under the infinitesimal time displacement $\xi^0$: any background scalar $X(\eta)$ obeys $\widetilde{\delta X}=\delta X-X'\xi^0$.  The spatial-curvature potential receives the compensating shift $\mathcal H\xi^0$.  Substitution shows explicitly that the combinations $\zeta$, $\zeta_n$, and $S$ are invariant \cite{Malik:2008im,Baumann:2009inflation}.}
Substitution immediately verifies that $\zeta$, $\zeta_n$ and $S$ are gauge invariant.  Physically, $S$ compares the local perturbation of the charge per comoving volume with the local perturbation of the total enthalpy.  An adiabatic perturbation corresponds to a local time delay along the homogeneous trajectory, for which
\begin{equation}
 \frac{\delta n}{n'}=\frac{\delta\rho}{\rho'}
 \quad\Longleftrightarrow\quad S=0.
 \label{eq:adiabaticcondition}
\end{equation}
Thus $S\neq0$ measures a displacement transverse to the background trajectory in thermodynamic state space.

Geometrically, an adiabatic perturbation shifts a local region along the homogeneous trajectory, whereas an entropy perturbation changes its composition transverse to that trajectory. Consequently, $S$ can remain nonzero even when the instantaneous total density perturbation vanishes.

Equation~\eqref{eq:Sdef} assumes exact conservation of the homogeneous charge.  If reactions violate the charge, $n'+3\cH n=a\mathcal C_n$, where $\mathcal C_n$ is the collision term per proper volume.  The gauge-invariant relative mode is still $3(\zeta_n-\zeta)$, but $n'$ must not be replaced by $-3\cH n$.  This distinction matters around chemical freeze-out, when $\mathcal C_n/Hn$ evolves through unity.

\subsection{Pressure decomposition and curvature sourcing}

For $p=p(\rho,n)$, the linear pressure perturbation can be decomposed as
\begin{equation}
 \delta p=c_a^2\delta\rho+\left(\frac{\partial p}{\partial n}\right)_\rho
 \left[\delta n-\frac{n'}{\rho'}\delta\rho\right],
 \qquad c_a^2\equiv\frac{p'}{\rho'}.
\end{equation}
 {Equation~(29) separates a pressure perturbation into a displacement along the homogeneous trajectory, $c_a^2\delta\rho$, and a composition perturbation transverse to that trajectory.  It follows from the total differential $\delta p=(\partial p/\partial\rho)_n\delta\rho+(\partial p/\partial n)_\rho\delta n$ after adding and subtracting $(\partial p/\partial n)_\rho(n'/\rho')\delta\rho$.  The bracket vanishes for a purely adiabatic time shift, so it isolates the entropy source \cite{Malik:2008im}.}
Using $n'/\rho'=n/(\rho+p)$ for separately conserved $n$, the nonadiabatic pressure is
\begin{equation}
 \delta p_{\rm nad}=\left(\frac{\partial p}{\partial n}\right)_\rho nS.
 \label{eq:pnadS}
\end{equation}
This identity gives a direct physical interpretation of mode conversion: a charge fluctuation gravitates as an entropy mode only if changing the charge at fixed total energy changes the pressure.  An exactly conformal equation of state $p=\rho/3$ has $(\partial p/\partial n)_\rho=0$ and therefore no super-Hubble conversion at linear order, even though $S$ itself may be nonzero.

The absence of conversion in the conformal limit is a statement about pressure response, not about the absence of charge fluctuations. A nonzero $S$ is present, but it cannot change $\zeta$ while the relation $p=\rho/3$ remains exact. Conversion begins only when a mass threshold, interaction correction, decay, or other nonconformal effect makes the pressure sensitive to composition at fixed energy density.

\section{Stochastic charge diffusion in an expanding universe}
\label{sec:diff}

The equilibrium covariance alone is not enough; transport determines which fluctuations survive. The physical wavenumber is $k_{\rm ph}=k/a$, where $k$ is the conserved comoving wavenumber. Diffusion relaxes shorter physical wavelengths faster because its rate scales as $Dk_{\rm ph}^2$.

\subsection{Constitutive relation and fluctuation-dissipation noise}

 {The hydrodynamic frame specifies how the local temperature, chemical potential, and velocity are defined away from exact equilibrium.  In the Landau frame the velocity is chosen so that the dissipative energy flux vanishes in the local rest frame, $u_\mu\delta T^{\mu\nu}=0$.  Charge may still diffuse relative to this energy flow, and that relative current is $\nu^\mu$.  This convention is standard in relativistic charged hydrodynamics \cite{Kovtun:2012rj,Kapusta:2012hydro}.}
For one $U(1)$ charge in the Landau frame,
\begin{equation}
 J^\mu=nu^\mu+\nu^\mu,
 \qquad u_\mu\nu^\mu=0,
\end{equation}
with first-order constitutive relation
\begin{equation}
 \nu^\mu=-\sigma T\Delta^{\mu\nu}\nabla_\nu\left(\frac{\mu}{T}\right)+I^\mu.
 \label{eq:constitutive}
\end{equation}
 {The deterministic part of Eq.~\eqref{eq:constitutive} is the relativistic form of Fick's law.  In an isothermal local rest frame, $\bm\nabla(\mu/T)=\bm\nabla\mu/T$ and $\delta n=\chi_{TT}\delta\mu$, so
\begin{equation*}
 \bm J_{\rm diff}=-\sigma\bm\nabla\mu=-\frac{\sigma}{\chi_{TT}}\bm\nabla n\equiv-D\bm\nabla n.
\end{equation*}
Fick's law states that the diffusive current points down the density gradient: particles migrate from regions of larger $n$ to regions of smaller $n$.  The coefficient $D>0$ has dimensions of length (or inverse energy in natural units) and sets the smoothing time of a physical Fourier mode, $\tau_{\rm diff}\simeq(Dk_{\rm ph}^2)^{-1}$.  The minus sign is required by positive entropy production, while $I^\mu$ restores the equilibrium fluctuations dissipated by the deterministic current \cite{Forster:1990hydro,Kovtun:2012rj,Kapusta:2012hydro}.}
Here $\sigma$ is the conductivity, $\Delta^{\mu\nu}=g^{\mu\nu}+u^\mu u^\nu$, and $I^\mu$ is stochastic noise.

The four-velocity $u^\mu$ satisfies $u^\mu u_\mu=-1$ in the metric convention used here. The projector $\Delta^{\mu\nu}$ removes the component parallel to the fluid velocity, so $\nu^\mu$ is a purely spatial dissipative current in the local rest frame. The conductivity $\sigma$ is nonnegative by entropy production. In local equilibrium its short-distance correlator is fixed by fluctuation-dissipation,
\begin{equation}
 \langle I^\mu(x)I^\nu(x')\rangle
 =2\sigma T\Delta^{\mu\nu}\frac{\delta^{(4)}(x-x')}{\sqrt{-g}},
 \label{eq:FDT}
\end{equation}
up to hydrodynamic-frame and regularization conventions.  The normalization follows from the local fluctuation-dissipation relation; analogous noise correlators in expanding relativistic fluids were developed explicitly in Ref.~\cite{Kapusta:2012hydro}. A causal theory replaces the white kernel by a colored kernel with finite current-relaxation time. Relativistic fluctuating hydrodynamics and its Schwinger-Keldysh formulation provide the systematic framework for this extension \cite{Kovtun:2012rj,Crossley:2017eft,Basar:2024qog,Mullins:2025crooks}.

When energy and momentum fluctuations can be neglected over the charge-relaxation interval, the linearized Fourier mode approximately obeys
\begin{equation}
 \dot{\delta n}_{\bm k}+3H\delta n_{\bm k}
 +D\frac{k^2}{a^2}\delta n_{\bm k}
 =\xi_{n,\bm k},
 \label{eq:langevin}
\end{equation}
 {Equation~\eqref{eq:langevin} follows by taking the covariant divergence of the current and linearizing about a homogeneous FLRW background.  The term $3H\delta n_{\bm k}$ dilutes a physical number density, the term $Dk^2/a^2$ damps spatial inhomogeneity according to Fick's law, and $\xi_{n,\bm k}$ is the divergence of the stochastic current.  The approximation neglects mixing with energy and momentum eigenmodes during the charge-relaxation interval \cite{Kovtun:2012rj,Kapusta:2012hydro}.}
where $D$ is the appropriate charge-diffusion eigenvalue.

The stochastic source $\xi_{n,\bm k}$ is the Fourier-space divergence of the current noise. The retarded kernel $G_k(t,t')$ subsequently introduced measures the survival of a fluctuation created at $t'$ until $t$; the factor $3H$ accounts for dilution of a physical number density, while $Dk^2/a^2$ describes genuine diffusive damping. In a multicomponent plasma, $D$ becomes a matrix built from conductivities and static susceptibilities. The formal unequal-time solution is
\begin{equation}
 \delta n_{\bm k}(t)=G_k(t,t_i)\delta n_{\bm k}(t_i)
 +\int_{t_i}^{t}\dd t'\,G_k(t,t')\xi_{n,\bm k}(t'),
 \label{eq:solution}
\end{equation}
 {Equation~\eqref{eq:solution} is obtained by the integrating-factor method.  The first term propagates an initial fluctuation, whereas the integral sums fluctuations injected continuously by the noise.  Their relative importance is fixed by fluctuation-dissipation: damping erases memory of the initial condition while noise repopulates the equilibrium variance \cite{Kapusta:2012hydro,Crossley:2017eft}.}
with
\begin{equation}
 G_k(t,t')=\exp\left[-\int_{t'}^t\dd\tau
 \left(3H+D\frac{k^2}{a^2}\right)\right].
\end{equation}
 {The retarded kernel is exponentially smaller than unity because both expansion and diffusion remove physical charge-density contrast.  A fluctuation created at $t'$ survives to $t$ only if the integrated dilution-plus-diffusion rate is not large.  For constant coefficients it reduces to $G_k=e^{-[3H+Dk^2/a^2](t-t')}$, making the two damping time scales explicit \cite{Forster:1990hydro,Kapusta:2012hydro}.}

\subsection{Diffusive freeze-out}
A mode follows the changing local-equilibrium distribution provided
\begin{equation}
 D\frac{k^2}{a^2}\gg H,
 \label{eq:equilibriumcondition}
\end{equation}
 {The physical meaning of Eq.~\eqref{eq:equilibriumcondition} is a comparison of clocks.  The mode relaxes toward its instantaneous equilibrium distribution on $\tau_{\rm diff}=(Dk^2/a^2)^{-1}$, whereas the background changes on $H^{-1}$.  When $\tau_{\rm diff}\ll H^{-1}$, many relaxation events occur in one expansion time and the mode adiabatically tracks equilibrium.  When the rates become comparable, tracking fails and the fluctuation freezes with a memory of the covariance near crossing \cite{Forster:1990hydro,Kovtun:2012rj}.}
while it freezes when
\begin{equation}
 D(T_k,\mu_k)\frac{k^2}{a_k^2}=c_D H_k,
 \qquad c_D=\order(1).
 \label{eq:freezeout}
\end{equation}
The corresponding physical diffusion length is
\begin{equation}
 \ell_D=\sqrt{\frac{D}{c_DH}}.
 \label{eq:ld}
\end{equation}
The equilibrium cell approximation requires
\begin{equation}
 \xi\ll\ell_D\ll H^{-1}.
 \label{eq:hierarchy}
\end{equation}
The second inequality is equivalent to $DH\ll1$ and allows charge fluctuations to freeze while the mode remains inside the Hubble radius.

The hierarchy $\xi\ll\ell_D$ ensures that each freeze-out region contains many approximately independent correlation cells, justifying Gaussian local equilibrium. The hierarchy $\ell_D\ll H^{-1}$ separates diffusive decoupling from Hubble crossing. Hence the conserved-charge calculation describes sub-Hubble freeze-out followed by later gravitational evolution. Later gravitational evolution then determines whether $S$ is conserved or converted into $\zeta$.

A useful sudden-freeze-out approximation replaces the full noise convolution by the equilibrium covariance at $T_k$ multiplied by a transfer matrix $\bm F(k)$. The approximation is reliable only if thermodynamic and transport quantities vary slowly across one relaxation time. The exact result is
\begin{equation}
 \mathcal P_{IJ}(k,t_f)=\frac{k^3}{2\pi^2}
 \int\dd t\,\dd t'\,
 G_{IA}(k;t_f,t)N_{AB}(k;t,t')G_{JB}(k;t_f,t'),
 \label{eq:master}
\end{equation}
where $N_{AB}$ is the energy-charge noise matrix. Equation~\eqref{eq:master}, rather than a single equal-time variance, is the appropriate starting point when $H\tau_{\rm rel}$ is not small.

\subsection{Diffusion constant and Einstein relation}

In the simplest single-charge problem, the conductivity and static susceptibility are related to the diffusion constant by the Einstein relation
\begin{equation}
 D=\frac{\sigma}{\chi_{TT}},
 \label{eq:Einsteinrelation}
\end{equation}
 {Equation~\eqref{eq:Einsteinrelation} is the Einstein relation.  It equates the Fick coefficient inferred from density-gradient transport with the conductivity multiplying the thermodynamic force $\nabla(\mu/T)$.  The susceptibility converts a density perturbation into its conjugate chemical-potential perturbation.  Thus a larger $\sigma$ accelerates diffusion, while a larger $\chi_{TT}$ stores more charge for the same chemical-potential gradient and slows the relaxation of $n$ \cite{Forster:1990hydro,Kovtun:2012rj}.}
with conventions in which the charge quantum is absorbed into $n$ and $\mu$.

Dimensionally, $D$ has units of inverse energy in natural units. The Einstein relation expresses the fact that a large static susceptibility reduces the chemical-potential gradient required to produce a given density gradient, while a large conductivity increases the corresponding current.  This follows by linearizing $n=n(T,\mu)$ at fixed temperature, so that $\bm\nabla(\mu/T)=\bm\nabla n/(T\chi_{TT})$, and comparing the constitutive current with Fick's law $\bm J=-D\bm\nabla n$.  For several charges, both conductivity and susceptibility are matrices and the diffusion operator is schematically $\bm D=\bm\sigma\bm\chi^{-1}$.  Its eigenvalues determine the relaxation rates.  Off-diagonal diffusion coefficients can be as large as diagonal entries in multicharge relativistic gases, so diagonalizing the transport problem is not optional in realistic baryon-electric-strangeness systems \cite{Greif:2018diffusion}.

Equation~\eqref{eq:langevin} is most transparent for the comoving fluctuation $\delta N_k=a^3\delta n_k$:
\begin{equation}
 \dot{\delta N}_{\bm k}+D\frac{k^2}{a^2}\delta N_{\bm k}
 =a^3\xi_{n,\bm k}.
 \label{eq:comovinglangevin}
\end{equation}
The expansion dilution term has disappeared.  Neglecting noise after a time $t_k$, the solution is suppressed by
\begin{equation}
 \mathcal D_k(t,t_k)=\exp[-k^2\mathcal I_D(t,t_k)],
 \qquad
 \mathcal I_D(t,t_k)=\int_{t_k}^{t}\frac{D(t')}{a^2(t')}\,\dd t'.
 \label{eq:dampingintegral}
\end{equation}
The comoving diffusion length is $k_D^{-2}=\mathcal I_D$; the physical diffusion length is $a\sqrt{\mathcal I_D}$.  The local criterion $Dk^2/a^2\simeq H$ follows when $D$, $a$, and $H$ vary by factors of order unity over one Hubble time.

\subsection{Causal correction}

First-order diffusion has dispersion relation $\omega=-iDk_{\rm ph}^2$ and infinite front velocity.  A minimal causal completion is the Maxwell-Cattaneo equation \cite{Cattaneo:1948,IsraelStewart:1979}
\begin{equation}
 \tau_J\Delta^\mu_{\ \nu}u^\alpha\nabla_\alpha\nu^\nu+\nu^\mu
 =-\sigma T\Delta^{\mu\nu}\nabla_\nu(\mu/T)+I^\mu,
 \label{eq:maxwellcattaneo}
\end{equation}
 {Equation~\eqref{eq:maxwellcattaneo} promotes the diffusion current to a relaxing degree of freedom.  Instead of responding instantaneously to a gradient, $\nu^\mu$ approaches the Navier-Stokes/Fick value over the microscopic time $\tau_J$.  This converts the parabolic diffusion equation, which has instantaneous tails, into a hyperbolic telegrapher equation with finite characteristic speed $\sqrt{D/\tau_J}$ \cite{Cattaneo:1948,IsraelStewart:1979,Jain:2024sk}.}
where $\tau_J$ is the current relaxation time.  In Minkowski space it gives
\begin{equation}
 \tau_J\omega^2+i\omega-Dk^2=0,
\end{equation}
and front speed $v_{\rm front}=\sqrt{D/\tau_J}$.  Causality requires $D/\tau_J\le1$ in units with $c=1$.

The relaxation time $\tau_J$ turns the parabolic diffusion equation into a hyperbolic telegrapher-type equation. The two roots of Eq.~\eqref{eq:maxwellcattaneo} contain a slowly relaxing diffusive branch and a rapidly damped transient branch. First-order hydrodynamics is recovered only at frequencies and wavenumbers well below $\tau_J^{-1}$.  The first-order freeze-out estimate is reliable when $H\tau_J\ll1$ and $k_{\rm ph}^2D\tau_J\ll1$.  Stable and causal Schwinger-Keldysh effective theories based on Maxwell-Cattaneo and Israel-Stewart dynamics provide a systematic treatment of the associated colored noise and higher-point functions \cite{Crossley:2017eft,Jain:2024sk,Mullins:2023info}.  In these formulations a local or dynamical KMS symmetry enforces fluctuation-dissipation and nonlinear Onsager constraints rather than imposing the noise kernel by hand \cite{Crossley:2017eft,Glorioso:2017kms}.

\subsection{Relation between diffusive freeze-out and horizon-scale matching}
\label{sec:two_matching_scales}
The conserved-charge calculation above and the open-subsystem calculation in Sec.~\ref{sec:open_curvature} use physically distinct matching scales. In the isolated plasma, the slow variable is a conserved charge density. Its relaxation rate is diffusive,
\begin{equation}
 \Gamma_n(k,t)=D(T,\mu)\frac{k^2}{a^2},
 \qquad \Gamma_n(k,t_k)=c_DH(t_k),
 \label{eq:charge_diffusive_rate}
\end{equation}
and the hierarchy $\ell_D\ll H^{-1}$ implies freeze-out while the mode is still sub-Hubble. The subsequent super-Hubble curvature is obtained only after evolving and projecting the frozen charge fluctuation.

By contrast, Secs.~\ref{sec:open_curvature} and \ref{sec:dark_fermion} describe the coarse-grained energy fluctuation of an open subsystem. Its use of $L=a/k\simeq H^{-1}$ is a horizon-scale matching prescription, not the diffusion condition in Eq.~\eqref{eq:charge_diffusive_rate}. More generally, an energy-like open-system mode has a relaxation rate $\Gamma_E(k,T,\mu)$ and freezes according to
\begin{equation}
 \Gamma_E(k,T_k,\mu_k)\simeq H_k.
 \label{eq:energy_freezeout_general}
\end{equation}
The identification $k=aH$ is justified only when the stochastic energy source remains in local equilibrium on sub-Hubble scales and the transition from local thermal fluctuations to a gravitationally constrained perturbation occurs over a Hubble time. The numerical example adopts this limit. It does not identify the open-sector energy mode with the conserved diffusive charge mode. A microscopic realization may instead possess a dissipative scale $k_F/a\neq H$; in that case Eqs.~\eqref{eq:zeta_open}-\eqref{eq:Pzeta_open} must be evaluated at that model-dependent scale.

\section{Primordial covariance at diffusive freeze-out}
\label{sec:spectra}

All spectra in this section are dimensionless unless a symbol $P_X(k)$ without calligraphic font is used. Specifically, $\mathcal P_X=k^3P_X/(2\pi^2)$. The subscript $k$ on a background quantity means evaluation at the mode-dependent diffusive freeze-out time fixed by Eq.~\eqref{eq:freezeout}.

In the Markovian sudden-freeze-out limit, the physical white-noise spectrum in Eq.~\eqref{eq:white} gives
\begin{equation}
 \mathcal P_{IJ}(k)
 =\left.\frac{k_{\rm ph}^3}{2\pi^2}
 \mathcal C_{IJ}(T,\mu)\right|_{k_{\rm ph}=k/a=\sqrt{c_DH/D}},
 \label{eq:pij}
\end{equation}
where $\mathcal C_{IJ}$ is defined by Eq.~\eqref{eq:projectedcov}. Thus
\begin{align}
 \PS(k)&=\left.\frac{1}{2\pi^2}
 \left(\frac{c_DH}{D}\right)^{3/2}
 \mathcal C_{SS}\right|_{T_k,\mu_k},
 \label{eq:PSgeneral}\\
 \mathcal P_{\zeta S}(k)&=\left.\frac{1}{2\pi^2}
 \left(\frac{c_DH}{D}\right)^{3/2}
 \mathcal C_{\zeta S}\right|_{T_k,\mu_k}.
 \label{eq:Pcrossgeneral}
\end{align}
The thermal correlation coefficient is
\begin{equation}
 \cos\Delta_{\rm th}(k)=
 \frac{\mathcal P_{\zeta S}}
 {\sqrt{\Pz\PS}}
 =\frac{\mathcal C_{\zeta S}}
 {\sqrt{\mathcal C_{\zeta\zeta}\mathcal C_{SS}}},
 \label{eq:correlationcoefficient}
\end{equation}
provided the two modes share the same freeze-out kernel.

If energy and charge relax with different kernels, the last equality in Eq.~\eqref{eq:correlationcoefficient} does not hold: unequal-time transport can rotate the covariance in fluctuation space. The equal-kernel expression should therefore be viewed as the controlled single-eigenmode limit. Positivity of $\Sigma$ guarantees $|\cos\Delta_{\rm th}|\le1$.

The spectral index of the isocurvature mode is
\begin{equation}
 n_{\rm iso}-1
 =\frac{\dd\ln\PS}{\dd\ln k}
 =\frac{
 \dd\ln\left[(H/D)^{3/2}\mathcal C_{SS}\right]/\dd\ln T
 }{
 \dd\ln\left[a(H/D)^{1/2}\right]/\dd\ln T
 }
 \bigg|_{T_k},
 \label{eq:tiltgeneral}
\end{equation}
where the background trajectory fixes $\mu(T)$. Equation~\eqref{eq:tiltgeneral} is the main model-independent tilt formula for diffusion-frozen thermal charge isocurvature.

The numerator measures how the freeze-out amplitude changes along the thermal background trajectory. The denominator converts temperature evolution into scale evolution because $k=a(H/D)^{1/2}$ at diffusive crossing. A nearly scale-invariant spectrum requires these two logarithmic slopes to nearly cancel.

\subsection{Fourier conventions and physical-to-comoving conversion}

We define the comoving Fourier transform by
\begin{equation}
 X(\bm x)=\int\frac{\dd^3k}{(2\pi)^3}X_{\bm k}e^{i\bm k\cdot\bm x},
 \qquad
 \langle X_{\bm k}X_{\bm k'}\rangle=(2\pi)^3\delta^{(3)}(\bm k+\bm k')P_X(k).
\end{equation}
The dimensionless spectrum is $\mathcal P_X=k^3P_X/(2\pi^2)$.  Since a physical wavevector is $\bm q=\bm k/a$ and a physical white-noise correlator is $P_{q_Aq_B}^{\rm ph}=\Sigma_{AB}$, the dimensionless spectrum at a fixed time is
\begin{equation}
 \mathcal P_{IJ}(k,t)=\frac{q^3}{2\pi^2}\mathcal C_{IJ}(t),
 \qquad q=\frac{k}{a}.
\end{equation}
This relation accounts for the factor $(H/D)^{3/2}$ in the freeze-out expression: the diffusion condition fixes $q_k=(c_DH/D)^{1/2}$, and an extensive thermal fluctuation is spatial white noise.  Window functions change the order-one amplitude but not this scaling.

\subsection{Sudden-freeze-out accuracy}

Let $\Gamma_k=Dk^2/a^2$ and define
\begin{equation}
 \epsilon_{\rm fr}\equiv\left|\frac{1}{\Gamma_k^2}\frac{\dd\Gamma_k}{\dd t}\right|_{t_k}.
\end{equation}
 {Equation~(56) is an adiabaticity parameter for freeze-out.  During one relaxation time $\Gamma_k^{-1}$, the fractional rate changes by approximately $|\dot\Gamma_k|/\Gamma_k^2$.  Therefore $\epsilon_{\rm fr}\ll1$ means that the relaxation rate is nearly constant while the mode equilibrates.  If it is order unity, the crossing is broad and the equal-time prescription has an order-one normalization uncertainty; the unequal-time kernel in Eq.~\eqref{eq:master} must then be integrated.}
The sudden approximation is parametrically controlled when $\epsilon_{\rm fr}\ll1$ and the equilibrium covariance changes slowly during a relaxation time,
\begin{equation}
 \epsilon_C\equiv\left|\Gamma_k^{-1}\frac{\dd\ln\mathcal C_{IJ}}{\dd t}\right|_{t_k}\ll1.
\end{equation}
At the nominal crossing $\Gamma_k\sim H$, these quantities are often order unity, so the amplitude carries a matching uncertainty.

This uncertainty affects the order-one normalization more strongly than the spectral slope. A reliable precision amplitude requires solving the unequal-time stochastic problem in Eq.~\eqref{eq:master}; the sudden prescription is best used to identify scaling laws and parametric dependence.  The spectral tilt remains more robust when $\epsilon_{\rm fr}$ and $\epsilon_C$ are approximately scale independent.  For precision predictions one should integrate Eq.~\eqref{eq:master}.  For white noise and slowly varying coefficients, the equal-time variance obeys a Lyapunov equation,
\begin{equation}
 \dot C_k=-2\Gamma_k(C_k-C_{k,\rm eq}),
 \label{eq:lyapunov}
\end{equation}
whose solution explicitly interpolates between equilibrium tracking and freeze-out.

\subsection{Amplitude estimate at diffusion crossing}

Take $\mathcal C_{SS}=A_ST^{-3}$, $H=1.66\sqrt{g_*}T^2/M_{\rm Pl}$, and $D=d_D/T$, the conformal weak- or strong-coupling scaling.  Equation~\eqref{eq:PSgeneral} gives
\begin{equation}
 \PS(T_k)=\frac{A_Sc_D^{3/2}}{2\pi^2d_D^{3/2}}
 \left(\sqrt{\frac{8\pi^3g_*}{90}}\frac{T_k}{M_{\rm Pl}}\right)^{3/2}.
 \label{eq:amplitudeestimate}
\end{equation}
Thus, even before imposing the blue tilt, the amplitude is suppressed by $(T/M_{\rm Pl})^{3/2}$ unless $A_S$ is enhanced by a small background asymmetry or proximity to a susceptibility peak.  For illustration, $g_*=100$, $d_D=c_D=A_S=1$, and $T_k=10^{10}\,\mathrm{GeV}$ give $\PS\sim10^{-13}$ up to reduced-versus-unreduced Planck-mass and matching conventions.  Raising the temperature increases the amplitude but also moves the perturbations toward shorter scales according to the freeze-out map.

This estimate displays the usual tension for an extensive thermal source. Raising $T_k$ weakens the Planck suppression, but it also moves freeze-out to a larger physical wavenumber. Enhancing the susceptibility through small charge yield, nonconformal dynamics, or critical correlations must be checked against linearity and the hierarchy in Eq.~\eqref{eq:hierarchy}.

\section{Conformal plasma and the universal blue spectrum}
\label{sec:conformal}

We now isolate the assumptions behind the blue spectrum. The conclusion $\mathcal P_S\propto k^3$ does not apply to every thermal system; it follows when the sector is conformal and extensive, $\mu/T$ is constant, the background is adiabatic, and the diffusion coefficient is a local power law.

\subsection{Equation of state and background evolution}
In this subsection, we consider the simplest analytic limit: an adiabatic, isolated conformal sector with no reservoir.
Consider a conformal plasma with one conserved charge,
\begin{equation}
 p(T,\mu)=T^4 f(x),
 \qquad x\equiv\frac{\mu}{T}.
 \label{eq:conformalp}
\end{equation}
Then
\begin{equation}
 \rho=3T^4f(x),
 \qquad n=T^3f'(x),
 \qquad \rho+p=4T^4f(x).
 \label{eq:conformalthermo}
\end{equation}
For adiabatic expansion with conserved entropy and charge, $n/s$ is constant. In a regular phase this fixes $x$ to a constant, so
\begin{equation}
 T\propto a^{-1},
 \qquad H\propto T^2.
 \label{eq:conformalbackground}
\end{equation}
 {By ``conformal susceptibilities'' we mean thermodynamic response functions in a scale-invariant plasma.  In $3+1$ spacetime dimensions, extensivity and the absence of an intrinsic mass scale imply $p=T^4f(\mu/T)$.  Each derivative with respect to the dimensionless charge source $\alpha=\mu/T$ changes only the function of $x=\mu/T$, whereas each derivative with respect to $-\beta$ introduces one additional power of $T$.  Consequently the charge, mixed, and energy covariance densities scale as $T^3$, $T^4$, and $T^5$.  Interactions may change the dimensionless functions of $x$ but not these powers as long as conformal invariance is exact.  Mass thresholds, running couplings, trace anomalies, and critical correlation lengths break this simple scaling \cite{Laine:2016thermal,Kovtun:2012rj}.}
Dimensional analysis of Eq.~\eqref{eq:SigmaDef} gives
\begin{equation}
 \Sigma_{\rho\rho}=T^5 A(x),
 \qquad
 \Sigma_{\rho n}=T^4 B(x),
 \qquad
 \Sigma_{nn}=T^3 C(x),
 \label{eq:conformalsigma}
\end{equation}
for dimensionless functions $A,B,C$.

The functions $A(x)$, $B(x)$, and $C(x)$ in Eq.~\eqref{eq:conformalsigma} are unrelated to the scalar metric perturbation $A$ or the later matching coefficient $\mathcal A$. Their arguments are the constant degeneracy parameter $x=\mu/T$, and their powers of $T$ follow solely from dimensional analysis in four spacetime dimensions. Every term in Eq.~\eqref{eq:CSS} therefore scales as
\begin{equation}
 \mathcal C_{SS}=T^{-3}\mathcal A_S(x).
 \label{eq:CSSscaling}
\end{equation}

\subsection{Power-law diffusion}

Let
\begin{equation}
 D(T)=D_0T^{-m},
 \label{eq:Dpower}
\end{equation}
with constant $m\neq0$. The freeze-out wavenumber is
\begin{equation}
 k_D(T)=a\sqrt{\frac{c_DH}{D}}
 \propto T^{-1}T^{(2+m)/2}=T^{m/2}.
 \label{eq:kDscaling}
\end{equation}
Hence
\begin{equation}
 T_k\propto k^{2/m}.
 \label{eq:Tk}
\end{equation}
Using Eqs.~\eqref{eq:PSgeneral}, \eqref{eq:CSSscaling}, and \eqref{eq:Dpower},
\begin{equation}
 \PS(k)
 \propto T_k^{\frac32(2+m)}T_k^{-3}
 =T_k^{3m/2}
 \propto k^3.
 \label{eq:k3}
\end{equation}
We therefore obtain
\begin{equation}
 n_{\rm iso}-1=3,\qquad n_{\rm iso}=4.
 \label{eq:niso4}
\end{equation}
The cancellation is independent of the power $m$.

Changing $m$ alters the map between wavelength and freeze-out temperature, but the equilibrium amplitude changes by precisely the compensating power. The resulting $k^3$ factor is therefore the dimensionless form of spatial white noise, not a special choice of transport microphysics. The case $m=0$ is degenerate: $k_D$ is constant during exact conformal radiation domination, so an extended range of modes does not successively freeze out.

Equation~\eqref{eq:niso4} is the principal analytic result. It shows that replacing Hubble crossing by diffusion crossing and introducing a finite chemical potential do not, by themselves, yield scale invariance. The result follows from three ingredients: conformal thermodynamics, extensive equilibrium fluctuations with finite correlation length, and power-law local diffusion along a trajectory of constant $\mu/T$.

\subsection{Conditions that modify the conformal result}

A spectrum different from Eq.~\eqref{eq:niso4} requires at least one underlying assumption to be relaxed. Useful possibilities are:
\begin{enumerate}
 \item \textit{Nonconformal susceptibility:} a mass threshold, phase transition, or strong trace anomaly changes the $T^{-3}$ scaling of $\mathcal C_{SS}$.
 \item \textit{Evolving $\mu/T$:} entropy production, charge transfer, or chemical freeze-out makes $x(T)$ nonconstant.
 \item \textit{Non-power-law transport:} critical slowing down or a sharp transport crossover makes $D(T)$ vary nonanalytically, with the freeze-out scaling controlled by the relevant dynamic universality class \cite{Hohenberg:1977dynamic,Son:2004qcd,DelCampo:2014kz}.
 \item \textit{Non-extensive fluctuations:} a correlation length comparable to the diffusion length invalidates Eq.~\eqref{eq:white}.
 \item \textit{Post-freeze-out conversion:} a scale-dependent transfer matrix can reshape the initial blue spectrum.
\end{enumerate}
Each possibility carries a corresponding consistency condition and can be assessed with the general expression in Eq.~\eqref{eq:tiltgeneral}.

\subsection{Direct derivation of the conformal scaling}

The scaling of the susceptibilities can be verified without leaving the source basis.  For $\Psi=VT^3f(x)$ and $x=\alpha$, differentiation at fixed $\alpha$ gives
 {Indeed, since $T=(-\lambda_E)^{-1}$ and $\partial_{(-\beta)}T=T^2$, one has
\begin{align*}
 \partial_{(-\beta)}\Psi
 &=T^2\partial_T[VT^3f(\alpha)]=3VT^4f(\alpha),\\
 \partial_{(-\beta)}^2\Psi
 &=T^2\partial_T[3VT^4f(\alpha)]=12VT^5f(\alpha),\\
 \partial_\alpha\partial_{(-\beta)}\Psi
 &=\partial_\alpha[3VT^4f(\alpha)]=3VT^4f'(\alpha),\\
 \partial_\alpha^2\Psi
 &=\partial_\alpha^2[VT^3f(\alpha)]=VT^3f''(\alpha).
\end{align*}
Dividing by $V$ gives Eqs.~(70)-(72).  The factors $12$, $3$, and $1$ are therefore consequences of the source derivatives, not model-dependent transport coefficients \cite{LandauLifshitz:1980statphys,Laine:2016thermal}.}
\begin{align}
 \Sigma_{\rho\rho}&=V^{-1}\partial_{(-\beta)}^2\Psi
 =12T^5f(x),\\
 \Sigma_{\rho n}&=V^{-1}\partial_{(-\beta)}\partial_\alpha\Psi
 =3T^4f'(x),\\
 \Sigma_{nn}&=V^{-1}\partial_\alpha^2\Psi
 =T^3f''(x).
 \label{eq:explicitconformalsigma}
\end{align}
The factors follow because $\partial_{(-\beta)}=T^2\partial_T$ at fixed $x$.  Substituting $n=T^3f'$, $\rho+p=4T^4f$ into Eq.~\eqref{eq:CSS} yields
\begin{equation}
 \mathcal C_{SS}=T^{-3}
 \left[\frac{3}{4f}-\frac{3}{2f}+\frac{f''}{f'^2}\right]
 =T^{-3}\left[\frac{f''}{f'^2}-\frac{3}{4f}\right].
 \label{eq:ASexplicit}
\end{equation}
Thermodynamic positivity ensures that the bracket is nonnegative.

The first term in the bracket is the normalized charge susceptibility, while the second subtracts the component aligned with the total energy fluctuation. Positivity states that the residual fluctuation orthogonal to the adiabatic direction has nonnegative variance.  The cross covariance is
\begin{equation}
\mathcal C_{\zeta S}=\frac{1}{12T^4f}[-3T+3T]=0.
 \label{eq:crossconformal}
\end{equation}
  The exact conformal equilibrium covariance therefore contains no curvature-isocurvature cross term in this basis. This is an equal-time local-equilibrium statement in the source basis. It does not imply that the final cosmological cross-spectrum must vanish: unequal relaxation kernels, nonconformal evolution, reservoir noise, or later entropy-to-curvature transfer can rotate the covariance and generate $\mathcal P_{\zeta S}\neq0$.  The fractional-charge variable still becomes singular as $f'\to0$, requiring a yield-based variable in a charge-symmetric background.

\subsection{Generalized power-law criterion}

The cancellation leading to Eq.~\eqref{eq:niso4} can be generalized.  Suppose along the background trajectory
\begin{equation}
 a\propto T^{-A},\qquad H\propto T^B,\qquad D\propto T^{-m},
 \qquad \mathcal C_{SS}\propto T^{-C}.
\end{equation}
 {Equation~(75) defines four local logarithmic slopes along the background trajectory: $A=-\dd\ln a/\dd\ln T$, $B=\dd\ln H/\dd\ln T$, $m=-\dd\ln D/\dd\ln T$, and $C=-\dd\ln\mathcal C_{SS}/\dd\ln T$.  They need not be global constants; over a sufficiently narrow temperature interval they can be interpreted as local slopes.  The formula that follows is meaningful only when $-A+(B+m)/2\neq0$, so that the freeze-out wavenumber changes monotonically with temperature.}
Then
\begin{equation}
 k_D\propto T^{-A+(B+m)/2},
 \qquad
 \PS\propto T^{\frac32(B+m)-C},
\end{equation}
and hence
\begin{equation}
 n_{\rm iso}-1=
 \frac{\frac32(B+m)-C}{-A+(B+m)/2}.
 \label{eq:generalpowercriterion}
\end{equation}
Exact scale invariance requires $C=3(B+m)/2$, while a nearly scale-invariant red spectrum requires a small negative numerator relative to the denominator.  For the conformal values $(A,B,C)=(1,2,3)$, Eq.~\eqref{eq:generalpowercriterion} reduces to $3$ for every $m\neq0$.  This formula is useful for mass thresholds and nonconformal dark sectors because $A$, $B$, $C$, and $m$ can be replaced by local logarithmic slopes.

\section{Open dark subsystem and thermal curvature perturbations}
\label{sec:open_curvature}

The open subsystem considered below is not a continuation of the conserved conformal calculation. Here the fluctuation scale is matched near Hubble crossing, the physical chemical potential is held approximately constant, and source terms maintain the dark component. The formulas below are therefore conditional on both the background exchange and the horizon-scale conversion prescription.

\subsection{Effective grand-canonical description}

Consider a relativistic dark fermion $X$ carrying a charge $Q_X$. During a finite interval, its reduced state is approximated by
\begin{equation}
 Z_X(\beta,\mu,V)
 =\operatorname{Tr}_X\exp\left[-\beta\left(H_X-\mu Q_X\right)\right],
 \qquad \beta=T^{-1},
 \label{eq:open_partition}
\end{equation}
where $H_X$ is the subsystem Hamiltonian, $V$ is a physical volume, and $\mu$ is an effective chemical potential. The approximation does not require the $X$ component to be isolated. Instead, the total stress tensor is conserved while the subsystem satisfies
\begin{align}
 \dot\rho_X+3H(\rho_X+p_X)=Q_E, \label{eq:open_continuity}\\
 \dot n_X+3Hn_X=Q_N.\label{eq:open_charge}
\end{align}
where $Q_E$ denotes energy transferred into $X$ per unit proper volume and proper time.

Similarly, $Q_N$ is the net charge transferred into the subsystem per unit proper volume and proper time. Positive $Q_E$ or $Q_N$ denotes injection into $X$. The reservoir carries $-Q_E$ and $-Q_N$ so that total energy-momentum and total charge remain conserved whenever the combined system has the corresponding symmetry. A reservoir carries the compensating source so that the total continuity equation remains homogeneous and covariantly conserved.

At the covariant level, the exchange is described by
\begin{equation}
 \nabla_\mu T_X^{\mu\nu}=Q_X^\nu,
 \qquad
 \nabla_\mu T_R^{\mu\nu}=-Q_X^\nu,
 \qquad
 Q_X^\nu=Q_Eu^\nu+F_X^\nu,
 \label{eq:covariant_exchange}
\end{equation}
where $u_\nu F_X^\nu=0$ and $F_X^\nu$ is the momentum-transfer four-vector. Charge exchange is similarly written as $\nabla_\mu J_X^\mu=Q_N$ with the compensating reservoir source. Equations~\eqref{eq:open_continuity} and \eqref{eq:open_charge} are the homogeneous limits of this covariant system. Gauge-invariant perturbations of interacting fluids, including energy and momentum transfer, are developed in Ref.~\cite{Malik:2002interacting}.

For the benchmark, the background exchange is specified phenomenologically by
\begin{equation}
 Q_E=(4-2\epsilon_H)H\rho_X,
 \qquad
 Q_N=Hn_X\,q_N(T,\mu),
 \label{eq:phenomenological_sources}
\end{equation}
where $q_N(T,\mu)$ is the dimensionless function given explicitly in Eq.~\eqref{eq:QN_qds}. This closure is sufficient to define the background trajectory but is not presented as a unique microscopic interaction. A complete realization must provide the perturbations $\delta Q_E$, $\delta Q_N$, and $F_X^\nu$ together with their noise kernels. Accordingly, the numerical spectrum is a conditional existence result for the specified sourced trajectory. Reservoir perturbations are not assumed to vanish in a fundamental model; their omission is part of the horizon-scale matching approximation encoded by $\mathcal A_X$.

Several mechanisms can motivate this effective description. The fermion may remain in chemical contact with heavier dark states, receive charge from a scalar condensate, interact with a slowly evolving homogeneous charge reservoir, or couple derivatively to a background field. For example,
\begin{equation}
 \mathcal L_{\rm int}=\frac{\partial_\mu\phi}{f}J_X^\mu
 \label{eq:derivative_coupling}
\end{equation}
produces an effective term $\dot\phi J_X^0/f$ in a homogeneous background and therefore an effective potential $\mu_{\rm eff}\simeq\dot\phi/f$. If $J_X^\mu$ is exactly conserved, this operator is a total derivative and cannot by itself generate $Q_N$. A viable realization based on Eq.~\eqref{eq:derivative_coupling} must therefore contain explicit charge transfer or charge violation, or else use a charged reservoir whose chemical equilibrium fixes the effective potential. The subsequent analysis does not select a particular microscopic realization. It assumes only that $|\dot\mu|\ll H|\mu|$ over the relevant interval and that local thermal equilibrium remains valid.

This open construction differs sharply from an isolated adiabatic plasma. If comoving charge and comoving entropy are separately conserved, the degeneracy parameter $x=\mu/T$ is approximately constant. Maintaining nearly constant $\mu$ while $T$ changes instead requires charge or energy exchange. This distinction is central to the relation between the blue conformal result in Sec.~\ref{sec:conformal} and the curvature spectrum derived below.

\subsection{Energy cumulants at finite chemical potential}

When $\ln Z$ is expressed in the variables $(\beta,\mu)$, energy cumulants are generated by differentiation at fixed $\alpha=\beta\mu$. It is convenient to introduce
\begin{equation}
 \mathfrak D
 \equiv-\left.\frac{\partial}{\partial\beta}\right|_\alpha
 =-\left.\frac{\partial}{\partial\beta}\right|_\mu
 +\frac{\mu}{\beta}\left.\frac{\partial}{\partial\mu}\right|_\beta.
 \label{eq:Doperator}
\end{equation}
The equality follows from $\mu=\alpha/\beta$.

The operator $\mathfrak D$ differentiates with respect to inverse temperature while preserving the dimensionless source $\alpha$. It therefore generates fluctuations of the physical energy $H_X$, rather than fluctuations of the grand-canonical combination $H_X-\mu Q_X$. This distinction is essential at finite chemical potential. In a cubic region of physical size $L$, the connected energy-density cumulants are
\begin{align}
 \rho&=\frac{1}{L^3}\mathfrak D\ln Z,\\
 \left\langle\delta\rho^2\right\rangle_L&=\frac{1}{L^3}\mathfrak D\rho,\\
 \left\langle\delta\rho^3\right\rangle_{c,L}&=\frac{1}{L^6}\mathfrak D^2\rho,\\
 \left\langle\delta\rho^4\right\rangle_{c,L}&=\frac{1}{L^9}\mathfrak D^3\rho.
 \label{eq:open_cumulants}
\end{align}
The powers of $L^{-3}$ reflect extensivity and the connected nature of the cumulants. Equation~\eqref{eq:Doperator} is the same source derivative that appears in the susceptibility formulation of Sec.~\ref{sec:thermo}, but it is now projected onto energy fluctuations rather than the relative charge mode.

The real-space variance is related to the mode amplitude by a window-dependent coefficient. For the Gaussian convention used here,
\begin{equation}
 \left|\delta\rho_k\right|^2
 =\frac{\gamma^2}{k^3}
 \left\langle\delta\rho^2\right\rangle_{L=a/k},
 \qquad
 \gamma=2\sqrt{2}\,\pi^{3/4}.
 \label{eq:window_gamma}
\end{equation}
Changing the window changes $\gamma$ but leaves the thermodynamic scaling and logarithmic slopes unchanged.

The coefficient $\gamma$ encodes only the normalization convention used to associate a real-space cell of size $L$ with a Fourier mode. Observable predictions should use one window convention consistently in the power spectrum and in all higher cumulants.

\subsection{Conversion to curvature perturbations}

Let
\begin{equation}
 \Omega_X\equiv\frac{\rho_X}{3\mpl^2H^2}
 =\frac{a^2\rho_X}{3\mpl^2\mathcal H^2}
 \label{eq:OmegaX}
\end{equation}
be the fractional background density carried by the fluctuating subsystem, where $H=\dot a/a$ is the physical Hubble parameter and $\mathcal H=aH$ is the conformal Hubble parameter. Around Hubble crossing, $L=a/k\simeq H^{-1}$, the gravitational constraint gives the transfer form, following the horizon-scale thermal matching strategy of Ref.~\cite{Biswas:2013lna},
\begin{equation}
 \zeta_k=\frac{\mathcal A(T_k)}{H_k^2\mpl^2}\delta\rho_k,
 \qquad
 \mathcal A(T)=\frac12\left[1+\frac{2(3+s_\rho)}{3(1+w_X)\Omega_X}\right].
 \label{eq:zeta_open}
\end{equation}
Here $w_X=p_X/\rho_X$ and
\begin{equation}
 s_\rho\equiv\frac{\dd\ln|\delta\rho|}{\dd\ln a}
 =-\frac32+\frac12\frac{\dd\ln(\mathfrak D\rho)}{\dd\ln a}
 \label{eq:s_open}
\end{equation}
measures the background scaling of the root-mean-square density fluctuation at fixed comoving scale. The first term in Eq.~\eqref{eq:s_open} comes from the physical volume $L^3=(a/k)^3$, and the second comes from the evolving grand-canonical susceptibility. The coefficient $\mathcal A$ should be regarded as a horizon-scale matching coefficient.

The large factor proportional to $\Omega_X^{-1}$ reflects the conversion of a fluctuation in a subdominant component into total curvature. It is not determined by equilibrium thermodynamics. When $\Omega_X$ is small, perturbations of the reservoir and the energy-transfer terms become especially important for verifying the matching prescription. A complete treatment would obtain it from the coupled gauge-invariant perturbation equations of the subsystem, reservoir, and dominant background.

Combining Eqs.~\eqref{eq:open_cumulants}, \eqref{eq:window_gamma}, and \eqref{eq:zeta_open} yields
\begin{equation}
 \mathcal P_\zeta(k)
 =\frac{\gamma^2}{2\pi^2}\mathcal A^2(T_k)
 \frac{\mathfrak D\rho_k}{H_k\mpl^4}
 =\frac{\gamma^2}{2\pi^2}\sqrt{3\Omega_X}\,\mathcal A^2(T_k)
 \frac{\mathfrak D\rho_k}{\mpl^3\sqrt{\rho_X}}.
 \label{eq:Pzeta_open}
\end{equation}
The factor $1/(2\pi^2)$ follows from the dimensionless-spectrum convention $\mathcal P_\zeta=k^3P_\zeta/(2\pi^2)$ and must be retained when converting the Fourier-mode variance in Eq.~\eqref{eq:window_gamma} to $\mathcal P_\zeta$.
 {The second form uses $H^2=\rho_X/(3\Omega_X\mpl^2)$.} The scalar tilt and running follow from
\begin{equation}
 n_s-1=\frac{\dd\ln\mathcal P_\zeta}{\dd\ln k},
 \qquad
 \alpha_s=\frac{\dd n_s}{\dd\ln k},
 \label{eq:tilt_open}
\end{equation}
with $T_k$ determined from $k=a(T_k)H(T_k)$.

For comparison with the standard observational convention, let $\mathcal P_t$ denote the primordial tensor power summed over the two helicities. Assigning the usual vacuum initial state gives the standard two-helicity spectrum~\cite{Baumann:2009inflation}
\begin{equation}
 \mathcal P_t(k)=\frac{2H_k^2}{\pi^2\mpl^2}
 =\frac{2\rho_k}{3\pi^2\mpl^4\Omega_{X,k}}.
 \label{eq:Ph_open}
\end{equation}
Combining this expression with Eq.~\eqref{eq:Pzeta_open}, the    tensor-to-scalar ratio is
\begin{equation}
 r_{t/s}(k)\equiv\frac{\mathcal P_t(k)}{\mathcal P_\zeta(k)}
 =\frac{4}{3\sqrt{3}\gamma^2}
 \frac{\rho_k^{3/2}}
 {\Omega_{X,k}^{3/2}\mathcal A_k^2\mpl\,\mathfrak D\rho_k}.
 \label{eq:r_open}
\end{equation}

We define windowed intrinsic local cumulant amplitudes by matching to the standard local expansion~\cite{KomatsuSpergel:2001local}
\begin{equation}
 \zeta=\zeta_g+\frac{3}{5}f_{\rm NL}^{}(\zeta_g^2-\langle\zeta_g^2\rangle)
 +\frac{9}{25}g_{\rm NL}^{}\zeta_g^3+\cdots .
\end{equation}
Thus $\langle\zeta_R^3\rangle_c=(18/5)f_{\rm NL}^{}\langle\zeta_R^2\rangle^2$ and the intrinsic contact part obeys $\langle\zeta_R^4\rangle_{c,{\rm contact}}=(216/25)g_{\rm NL}^{}\langle\zeta_R^2\rangle^3$.  The $f_{\rm NL}^2$ exchange contribution to the full trispectrum is separate.  With the same Gaussian window and horizon matching as the scalar spectrum,
\begin{align}
 f_{\rm NL}^{}(k)&=
 \frac{5}{24\gamma\mathcal A(T_k)\Omega_{X,k}}
 \frac{\rho\,\mathfrak D^2\rho}{(\mathfrak D\rho)^2},
 \label{eq:fNL_open}\\
 g_{\rm NL}^{}(k)&=
\frac{25}{486\gamma^2\mathcal A^2(T_k)\Omega_{X,k}^2}
 \frac{\rho^2\mathfrak D^3\rho}{(\mathfrak D\rho)^3}.
 \label{eq:gNL_open}
\end{align}

\section{Relativistic dark-fermion example}
\label{sec:dark_fermion}

For a massless Dirac fermion, all thermodynamic quantities needed above are analytic. The symbol $X$ labels the dark fermion sector, $R$ labels the reservoir, $N=\ln(a/a_\star)$ counts e-folds from pivot exit, and a star denotes evaluation when $k_\star=a_\star H_\star$. The parameter $\epsilon_H=-\dot H/H^2$ is constant and lies between zero and one during accelerated expansion.

\subsection{Equation of state and quasi-de Sitter closure}

For one effectively massless Dirac fermion,
\begin{align}
 \rho_X(T,\mu)&=\frac{7\pi^2}{60}T^4+\frac12\mu^2T^2
 +\frac{\mu^4}{4\pi^2},
 &p_X&=\frac13\rho_X,\label{eq:MasslessFermion}\\
 n_X(T,\mu)&=\frac13\mu T^2+\frac{\mu^3}{3\pi^2}.
 \label{eq:fermion_number}
\end{align}
Equation~\eqref{eq:Doperator} then yields
\begin{equation}
 \left\langle\delta\rho^2\right\rangle=4T\rho_X,
 \qquad  \left\langle\delta\rho^3\right\rangle=20T^2\rho_X,
 \qquad  \left\langle\delta\rho^4\right\rangle=120T^3\rho_X.
 \label{eq:fermion_cumulants}
\end{equation}

Here $\mu/T$ measures the importance of the charge asymmetry in the Fermi-Dirac distributions. The regime $|x|=\order(1)$ interpolates smoothly between small asymmetry and strong degeneracy: the pressure is analytic for $T>0$, and $\chi_{TT}=T^2/3+\mu^2/\pi^2$ is finite and positive \cite{Laine:2016thermal}. In particular, $T\sim|\mu|$ is not a phase-transition criterion for this gas.
The three contributions to $\rho_X$ are the purely thermal term, the mixed thermal-density term, and the zero-temperature degenerate term. Their smooth interpolation and the positivity of $\chi_{TT}$ show that no critical enhancement is present in the free massless gas.

We assume a finite interval with constant $0<\epsilon_H<1$ and constant physical $\mu$.  With $N=0$ at the pivot exit event,
\begin{equation}
 H(N)=H_\star e^{-\epsilon_HN},
 \qquad \frac{k}{k_\star}=e^{(1-\epsilon_H)N},
 \qquad T_\star=T(0).
 \label{eq:qds_background}
\end{equation}
The second relation follows from evaluating each mode at $k=aH$.  The energy fraction $\Omega_X=\rho_X/(3\mpl^2H^2)$ is generally time dependent: Eq.~\eqref{eq:open_continuity} gives the exact background identity
\begin{equation}
 \frac{\dd\ln\Omega_X}{\dd N}
 =-4+\frac{Q_E}{H\rho_X}+2\epsilon_H.
 \label{eq:fermion_fraction_evolution}
\end{equation}
In particular, without energy exchange it decreases as
$\Omega_X\propto e^{-(4-2\epsilon_H)N}$.  For the example below we impose a constant $0<\Omega_X<1$.  This is a sourced tracking assumption: $\rho_X$ follows $H^2$, and the required energy supply is given below.

Constant $\Omega_X$ means that the dark component redshifts at the same fractional rate as the total background. Since a freely redshifting relativistic gas would dilute as $a^{-4}$, the source must replenish almost four Hubble-dilution units of energy per e-fold. Equation~\eqref{eq:QE_qds} quantifies this statement.
\begin{equation}
 \frac{Q_E}{H\rho_X}=4-2\epsilon_H.
 \label{eq:QE_qds}
\end{equation}
It compensates most of the dilution that an isolated radiation component would experience.  The background assumptions specify this exchange rate; they do not derive it from a microscopic interaction.

At fixed $\mu$, the temperature trajectory is determined directly by the energy density,
\begin{equation}
 \frac{7\pi^2}{60}T^4(N)+\frac12\mu^2T^2(N)+\frac{\mu^4}{4\pi^2}
 =3\Omega_X\mpl^2H_\star^2 e^{-2\epsilon_HN}.
 \label{eq:fermion_temperature_trajectory}
\end{equation}
The positive-temperature branch can be written explicitly as
\begin{equation}
 T^2(N)=\frac{30}{7\pi^2}
 \left[\sqrt{\frac{2\mu^4}{15}
 +\frac{7\pi^2}{5}\Omega_X\mpl^2H^2(N)}-\frac{\mu^2}{2}\right].
 \label{eq:fermion_temperature_solution}
\end{equation}

Differentiating Eq.~\eqref{eq:fermion_temperature_trajectory} at fixed physical $\mu$ gives
\begin{equation}
 \frac{\dd\ln T}{\dd N}
 =-2\epsilon_H
 \frac{\frac{7\pi^2}{60}+\frac{\mu^2}{2T^2}+\frac{\mu^4}{4\pi^2T^4}}
 {\frac{7\pi^2}{15}+\frac{\mu^2}{T^2}}.
 \label{eq:fermion_cooling}
\end{equation}
For $\mu\ne0$, $|\mu|/T$ grows as the subsystem cools, even though both $\mu$ and $\Omega_X$ are held fixed.  Using this cooling rate in Eq.~\eqref{eq:open_charge} fixes the charge exchange as well:
\begin{equation}
 \frac{Q_N}{Hn_X}
 =3-4\epsilon_H
 \frac{\frac{7\pi^2}{60}+\frac{\mu^2}{2T^2}+\frac{\mu^4}{4\pi^2T^4}}
 {\left(\frac{7\pi^2}{15}+\frac{\mu^2}{T^2}\right)
  \left(1+\frac{\mu^2}{\pi^2T^2}\right)}.
 \label{eq:QN_qds}
\end{equation}
This ratio is defined for $\mu\ne0$; at $\mu=0$ the net charge density and its required source both vanish.

Assuming that $X$ and a reservoir $R$ exhaust the total background density, the Friedmann equations fix the reservoir pressure to be $p_R=w_R\rho_R$, with
\begin{equation}
 w_R=-1+\frac{2\epsilon_H-4\Omega_X}{3(1-\Omega_X)}.
 \label{eq:w_reservoir}
\end{equation}
For positive $\rho_R$, its null energy condition $w_R\ge-1$ is equivalent to $\Omega_X\le\epsilon_H/2$.  The same condition follows from the total enthalpy balance $2\epsilon_H\mpl^2H^2=4\rho_X/3+(\rho_R+p_R)$.

The balance $Q_E\simeq4H\rho_X$ is analogous to the replenishment of radiation in warm inflation. Explicit constructions, including the Warm Little Inflaton with light fermions, show how such a bath can coexist with accelerated expansion \cite{BasteroGil:2016warm}.
However, warm-inflation calculations evolve coupled inflaton, radiation and metric fluctuations, and their dissipative freeze-out scale need not coincide with $k=aH$ \cite{Hall:2004warm}.

\subsection{Spectrum and numerical example}

The matching prescription of section~\ref{sec:open_curvature} now has a fully specified background.  Substituting Eq.~\eqref{eq:fermion_cooling} into Eqs.~\eqref{eq:zeta_open} and \eqref{eq:s_open} gives the coefficient already defined there,
\begin{equation}
 \mathcal A_X
 =\frac{1}{8\Omega_X}\left[
 4\Omega_X+3-2\epsilon_H
 -2\epsilon_H
 \frac{\frac{7\pi^2}{60}+\frac{\mu^2}{2T^2}+\frac{\mu^4}{4\pi^2T^4}}
 {\frac{7\pi^2}{15}+\frac{\mu^2}{T^2}}\right].
 \label{eq:A_qds}
\end{equation}
Writing the spectrum with this coefficient substituted explicitly,
\begin{align}
 \mathcal P_{\zeta_X}(k)
 ={}&\frac{\gamma^2T^5}{32\pi^2\Omega_X^2H\mpl^4}
 \left(\frac{7\pi^2}{60}+\frac{\mu^2}{2T^2}
 +\frac{\mu^4}{4\pi^2T^4}\right)\nonumber\\
 &\times\left[
 4\Omega_X+3-2\epsilon_H-2\epsilon_H
 \frac{\frac{7\pi^2}{60}+\frac{\mu^2}{2T^2}+\frac{\mu^4}{4\pi^2T^4}}
 {\frac{7\pi^2}{15}+\frac{\mu^2}{T^2}}\right]^2.
 \label{eq:fermion_spectrum_explicit}
\end{align}
Here $T$ and $H$ are evaluated at the exit time of $k$ using Eqs.~\eqref{eq:qds_background} and \eqref{eq:fermion_temperature_solution}.

 Differentiating with $\dd/\dd\ln k=(1-\epsilon_H)^{-1}\dd/\dd N$ gives
\begin{align}
 n_s-1={}&-\frac{\epsilon_H}{1-\epsilon_H}
 \left[1+2
 \frac{\frac{7\pi^2}{60}+\frac{\mu^2}{2T^2}+\frac{\mu^4}{4\pi^2T^4}}
 {\frac{7\pi^2}{15}+\frac{\mu^2}{T^2}}\right]\nonumber\\
 &-\frac{16\epsilon_H^2}{1-\epsilon_H}\frac{\mu^2}{T^2}
 \frac{\left(\tfrac{7\pi^2}{60}+\tfrac{\mu^2}{2T^2}+\tfrac{\mu^4}{4\pi^2T^4}\right)
       \left(\tfrac{7\pi^2}{60}+\tfrac{7\mu^2}{30T^2}+\tfrac{\mu^4}{4\pi^2T^4}\right)}
 {\left(\tfrac{7\pi^2}{15}+\tfrac{\mu^2}{T^2}\right)^2
  \left[\tfrac{7\pi^2}{60}(16\Omega_X+12-10\epsilon_H)
       +(4\Omega_X+3-3\epsilon_H)\tfrac{\mu^2}{T^2}
       -\tfrac{\epsilon_H\mu^4}{2\pi^2T^4}\right]}.
 \label{eq:qds_tilt}
\end{align}
The first line accounts for the cooling and the evolution of $H$; the second is the contribution from the changing matching coefficient.

The first contribution is present even at $\mu=0$. The second is proportional to $\mu^2/T^2$ and therefore isolates the additional scale dependence produced by finite chemical potential through the evolution of $\mathcal A_X$. This decomposition explains why finite $\mu$ modifies the tilt but is not required for a red spectrum on the sourced trajectory.

The benchmark uses the following assumptions: constant $\epsilon_H$, constant $\Omega_X$, constant physical $\mu$, local grand-canonical equilibrium, horizon-scale matching $L\simeq H^{-1}$ for the open energy mode, a Gaussian smoothing convention, and the effective coefficient $\mathcal A_X$. Reservoir and source perturbations are not independently evolved. These assumptions define the scope of the existence proof and separate it from the sub-Hubble diffusive charge calculation.

Our numerical example looks for parameters $(\mu/T_\star,\Omega_X)$. We solve for $n_s(k_\star)=0.9649$ from $\epsilon_H$, from $\mathcal P_{\zeta_X}^{}(k_\star)=2.10\times10^{-9}$ we fix $T_\star/\mpl$, using $k_\star=0.05\,\mathrm{Mpc}^{-1}$; these are the Planck pivot targets \cite{Planck:2018inflation}.

 A root in $\epsilon_H$ fixes the pivot tilt, and the amplitude fixes $T_\star/\mpl$ algebraically because, at fixed $\mu/T_\star$, $\Omega_X$ and $\epsilon_H$, Eq.~\eqref{eq:fermion_spectrum_explicit} scales as $\mathcal P_{\zeta_X}\propto(T_\star/\mpl)^3$.
 Fixing $\mu/T_\star$ specifies a pivot input across candidate models; within each model it is the physical $\mu$, rather than the evolving ratio $\mu/T$, that remains constant.
 The representative choice $\mu/T_\star=2$, $\Omega_X=0.003$ gives the values in Table~\ref{tab:benchmark}.

These two quantities are benchmark inputs rather than fitted cosmological posteriors. At fixed values of them, $\epsilon_H$ is chosen to reproduce the pivot tilt and $T_\star/\mpl$ is then chosen to reproduce the pivot amplitude. All source rates, hierarchy ratios, and higher cumulants are outputs of that     construction.

\begin{table}[tbp]
\centering
\caption{Representative     pivot solution for $\mu/T_\star=2$ and $\Omega_X=0.003$. The parameters $\epsilon_H$ and $T_\star/\mpl$ are fixed by the target scalar tilt and amplitude at $k_\star=0.05\,\mathrm{Mpc}^{-1}$. The remaining entries follow from the sourced constant-$\Omega_X$ background and the horizon-scale matching prescription.}
\label{tab:benchmark}
\begin{tabular}{@{}ll@{}}
\toprule
Quantity & Value \\
\midrule
$\mu/T_\star$ & $2$ \\
$\Omega_X$ & $0.003$ \\
$\epsilon_H$ & $0.0187968$ \\
$T_\star/\mpl$ & $4.40621\times10^{-5}$ \\
$\mu/\mpl$ & $8.81244\times10^{-5}$ \\
$H_\star/\mpl$ & $3.85955\times10^{-8}$ \\
$T_\star/H_\star$ & $1141.64$ \\
$\mathcal A_{X\star}$ & $123.286$ \\
$\alpha_s(k_\star)$ & $-1.66951\times10^{-4}$ \\
$Q_E/(H\rho_X)$ & $3.96241$ \\
$Q_N/(Hn_X)$ & $2.97789$ \\
$w_R$ & $-0.991443$ \\
\bottomrule
\end{tabular}
\end{table}
Table~\ref{tab:benchmark} shows a clear hierarchy of scales: $T_\star/H_\star\simeq1142$ places the local fermion bath well above the expansion rate, while $T_\star$ and $H_\star$ remain sub-Planckian. The source terms are substantial, $Q_E/(H\rho_X)\simeq3.96$ and $Q_N/(Hn_X)\simeq2.98$, confirming that constant $\Omega_X$ and constant physical $\mu$ require continuous energy and charge exchange. The value $w_R\simeq-0.991$ supports accelerated expansion and satisfies $\Omega_X\leq\epsilon_H/2$ for the benchmark.

The temperature in Table~\ref{tab:benchmark} is the local matter temperature, clearly distinct from the Gibbons-Hawking temperature $T_{\rm dS}=H/(2\pi)$ \cite{GibbonsHawking:1977}. At the pivot, $T_\star/T_{{\rm dS},\star}\simeq2.02\times10^3$. The example therefore describes a hot, sourced subsystem during quasi-de Sitter expansion. Horizon thermality alone does not supply the assumed thermal bath, charge asymmetry or exchange rates.

\begin{figure}[tbp]
    \centering
    \includegraphics[width=0.91\linewidth]{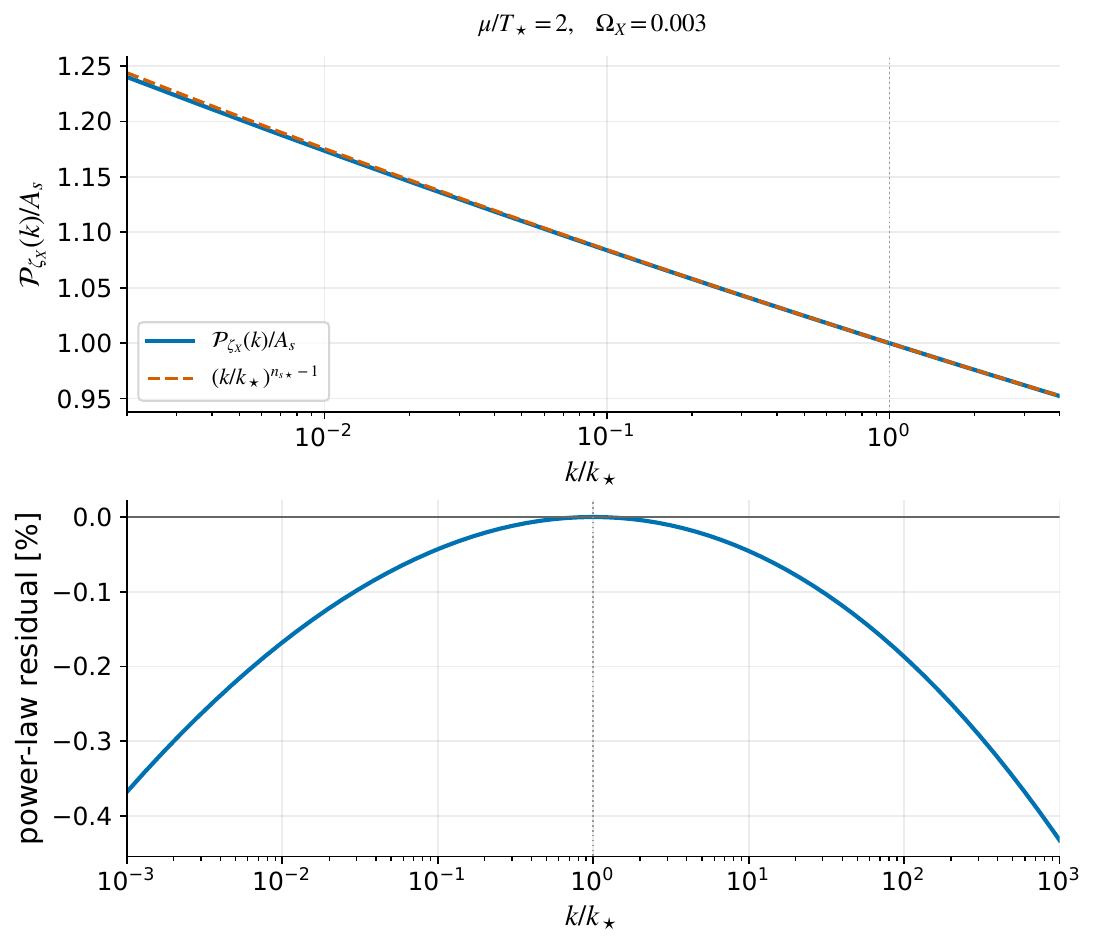}
    \caption{Curvature spectrum for the representative solution $\mu/T_\star=2$ and $\Omega_X=0.003$. The upper panel compares $\mathcal P_{\zeta_X}(k)/A_s$ (solid blue curve) with the pivot power law $(k/k_\star)^{n_{s\star}-1}$ (black dashed curve), using $A_s=2.10\times10^{-9}$ and $n_{s\star}=0.9649$. The lower panel shows the fractional departure $100[\mathcal P_{\zeta_X}/(A_s(k/k_\star)^{n_{s\star}-1})-1]$ over $10^{-3}\leq k/k_\star\leq10^3$. Within the fitted interval $0.002\leq k/k_\star\leq4$, the maximum absolute residual is $3.003\times10^{-3}$, or approximately $0.30\%$, demonstrating that the thermal spectrum is accurately described by a weakly running red power law across the range used in the numerical fit.}
    \label{fig:pzeta}
\end{figure}

\begin{figure}[tbp]
    \centering
    \includegraphics[width=0.7\linewidth]{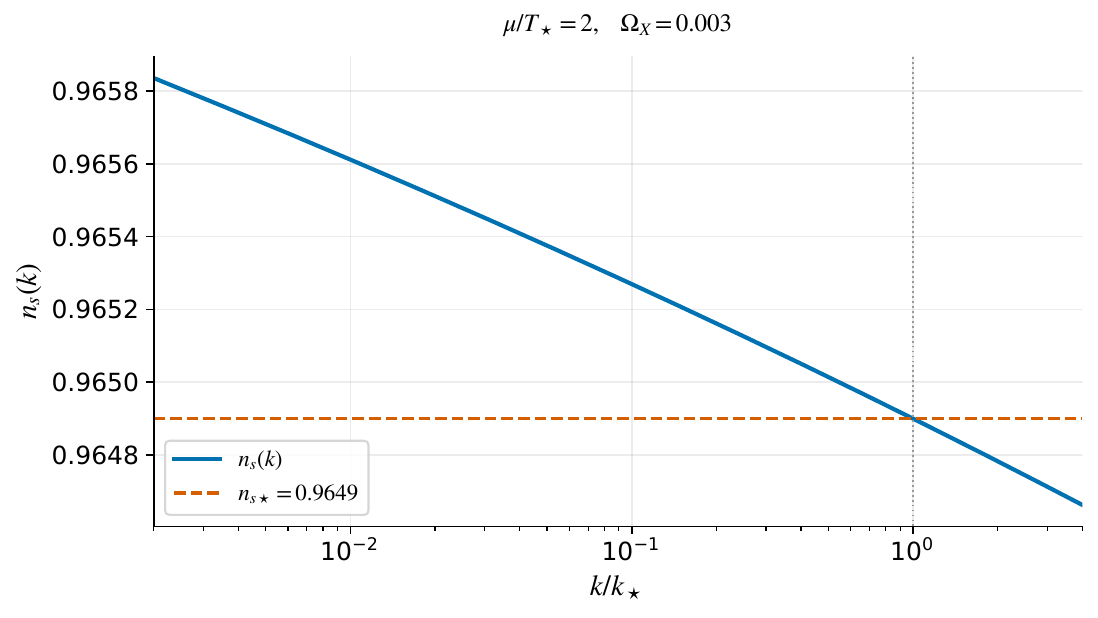}
    \caption{Scale dependence of the local scalar index $n_s(k)=1+\dd\ln\mathcal P_{\zeta_X}/\dd\ln k$ for $\mu/T_\star=2$ and $\Omega_X=0.003$. The solid blue curve is the model prediction and the black dashed line marks the pivot value $n_{s\star}=0.9649$. The index changes only at the level of a few parts in $10^{-4}$ over the displayed fitted interval, consistently with the small running $\alpha_s(k_\star)=-1.66951\times10^{-4}$.}
    \label{fig:spectral_diagnostics}
\end{figure}

Across $0.002\le k/k_\star\le4$, the spectrum differs from the pivot power law by at most $3.003\times10^{-3}$; an unweighted least-squares fit of $\ln\mathcal P$ against $\ln k$ on the logarithmic output grid gives $n_s=0.965279$.  Figures~\ref{fig:pzeta} and \ref{fig:spectral_diagnostics} display the spectrum, its residual relative to the pivot power law, and the corresponding local spectral index.  This benchmark is not unique. In particular, at $\mu=0$ Eq.~\eqref{eq:fermion_cooling} gives $\dd\ln T/\dd N=-\epsilon_H/2$, while Eq.~\eqref{eq:A_qds} gives the constant coefficient $\mathcal A_X=1/2+(3-5\epsilon_H/2)/(8\Omega_X)$.  Consequently,
\begin{equation}
 n_s-1=-\frac{3\epsilon_H}{2(1-\epsilon_H)},\qquad
 \epsilon_H=\frac{1-n_s}{5/2-n_s}=0.02286496
 \quad\hbox{for }n_s=0.9649.
 \label{eq:zero_mu_qds}
\end{equation}
The amplitude can again be fitted by $T_\star$.  Thus finite chemical potential is not necessary for this red     spectrum; the sourced constant-fraction accelerating trajectory already suffices.

Figure~\ref{fig:pzeta} shows the main numerical comparison. The sourced quasi-de Sitter trajectory replaces the strongly blue behavior of the isolated conformal diffusion channel with a spectrum that closely follows a red power law. The smooth residual indicates that the departure from a pure power law is generated by the slow evolution of $T/H$ and $\mathcal A_X$, rather than by a sharp transition. The larger departure outside the fitted interval is not an observable-scale prediction unless the duration of the generating phase and the later expansion history are specified.

Figure~\ref{fig:spectral_diagnostics} shows that the local index remains close to the pivot target throughout the fitted range. Together with Fig.~\ref{fig:pzeta}, this demonstrates that the agreement is not confined to a single scale. The quoted running, $\alpha_s(k_\star)=-1.66951\times10^{-4}$, is sufficiently small that only a weak accumulated departure from the pivot power law develops over the sampled interval.

For the representative point, the standard two-helicity normalization in Eq.~\eqref{eq:r_open} gives
\begin{equation}
 r_{t/s}(k_\star)=1.82\times10^{-4}.
 \label{eq:r_benchmark}
\end{equation}
This value is obtained by the exact factor-of-eight normalization conversion and does not require a new numerical background calculation. It remains small, but its interpretation is based on the assumed vacuum tensor state.

The corrected benchmark value in Eq.~\eqref{eq:r_benchmark} is stated in the standard two-helicity convention.

\begin{figure}[tbp]
    \centering
    \includegraphics[width=0.7\linewidth]{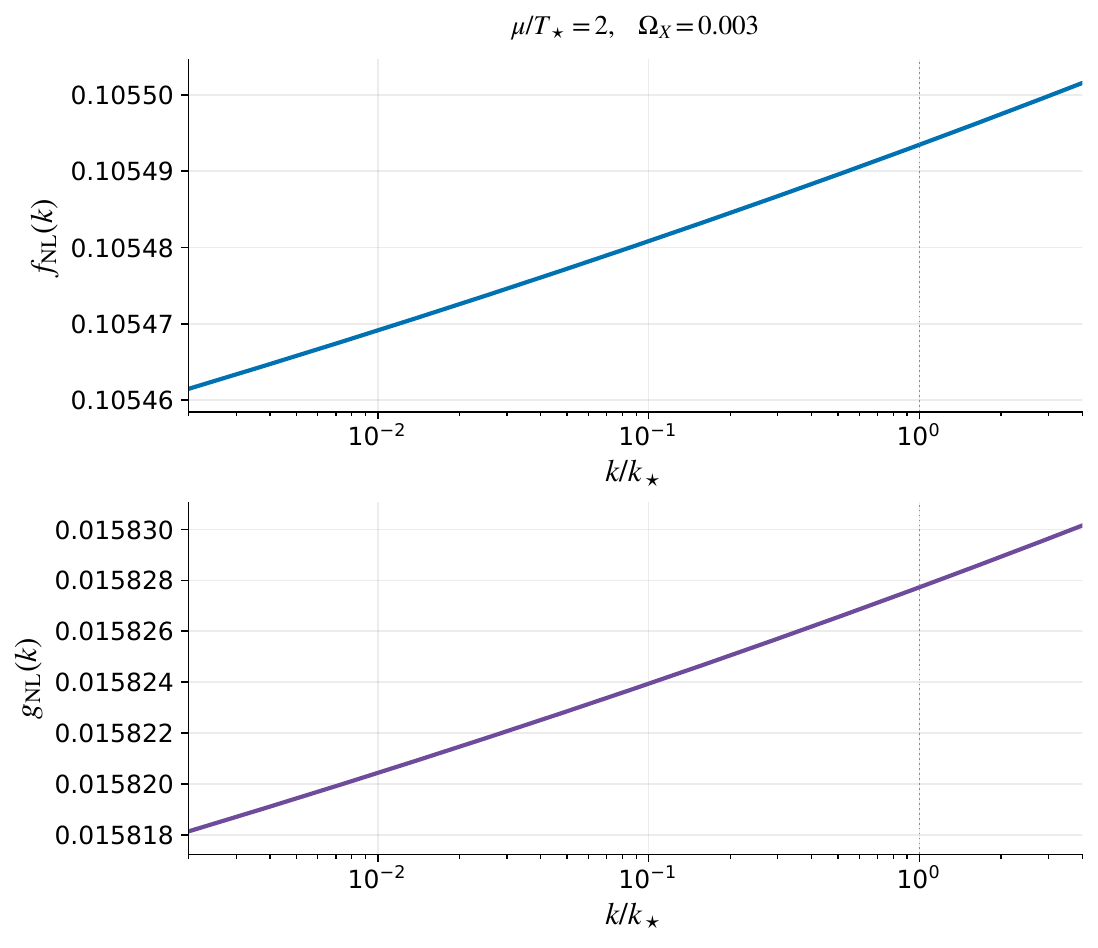}
    \caption{ {Scale dependence of the local-type cumulant amplitudes for $\mu/T_\star=2$ and $\Omega_X=0.003$. The upper panel shows $f_{\rm NL}^{}(k)$ and the lower panel shows $g_{\rm NL}^{}(k)$, evaluated using the grand-canonical energy cumulants and the horizon-scale matching prescription. At the pivot, $f_{\rm NL}^{}=0.1055$ and $g_{\rm NL}^{}=0.01583$. Both quantities vary by less than approximately $10^{-3}$ fractionally over the fitted interval.}}
    \label{fig:ng}
\end{figure}

Eqs.~\eqref{eq:fermion_cumulants}  give
\begin{equation}
 f_{\rm NL}^{}=\frac{25}{96\gamma\Omega_X\mathcal A_X},
 \qquad
 g_{\rm NL}^{}=\frac{125}{1296\gamma^2\Omega_X^2\mathcal A_X^2},
 \label{eq:contact_fermion}
\end{equation}
which evaluate to
\begin{equation}
     f_{\rm NL}^{}(k_\star)=0.1055, \qquad g_{\rm NL}^{}(k_\star)=0.01583
\end{equation}
at the pivot. Figure~\ref{fig:ng} displays the scale dependence of $f_{\rm NL}$ and $g_{\rm NL}$; the corrected tensor ratio is given analytically in Eq.~\eqref{eq:r_benchmark}.

Figure~\ref{fig:ng} shows that the third- and fourth-order grand-canonical cumulants give small positive values of $f_{\rm NL}$ and $g_{\rm NL}$ with negligible scale dependence across the fitted interval. The near constancy follows from fixed $\Omega_X$ and slowly varying $\mathcal A_X$. These amplitudes use a local cumulant normalization; comparison with CMB bispectrum and trispectrum templates additionally requires their momentum dependence and transfer through the reservoir and radiation sectors. Related warm-inflation analyses likewise show that dissipation and radiation noise can generate model-dependent higher-point structure beyond a single local amplitude \cite{BasteroGil:2014NG}.

Recent warm-inflation work makes the role of chemical potentials particularly relevant: axion-gauge models evolve chiral fermion asymmetries, while analyses of pseudoscalar couplings show that the bath's induced chemical potentials can modify the relation between effective inflaton friction and noise \cite{Berghaus:2025SM,Broadberry:2026chemical}. These are driven chemical responses, often associated with nonconserved charges, and cannot be identified directly with the thermodynamic charge potential used here. Likewise, chemical potentials that enhance particle production in cosmological-collider models need not describe an equilibrated gas or its statistical energy cumulants \cite{WangXianyu:2020chemical}.

For the derivative coupling in Eq.~\eqref{eq:derivative_coupling}, if $J_X^\mu$ is exactly conserved by all interactions, $(\partial_\mu\phi)J_X^\mu$ is a total derivative and cannot alone generate the required charge source. A realization based on this coupling must specify charge transfer or charge-violating dynamics and the resulting distribution, as in kinetic treatments of spontaneous baryogenesis \cite{Dasgupta:2018chemical}. A chemical potential maintained by a charged reservoir remains a distinct possibility.

\section{Super-Hubble evolution and mode conversion}
\label{sec:conversion}

Finally, we distinguish the covariance at freeze-out from the perturbations inherited by the later radiation era. The transfer coefficient $T_{\zeta S}$ measures entropy-to-curvature conversion, while $T_{SS}$ measures survival or damping of the entropy mode. Both may depend on $k$ if the transition history is scale dependent.

For multiple components, the total curvature perturbation evolves as
\begin{equation}
 \dot\zeta=-\frac{H}{\rho+p}\delta p_{\rm nad}
 -\frac{k^2}{3a^2}\mathcal V,
 \label{eq:zetadot}
\end{equation}
where $\delta p_{\rm nad}=\delta p-c_a^2\delta\rho$ and $\mathcal V$ is a gauge-invariant velocity potential. On super-Hubble scales the gradient term is negligible, but a charge isocurvature perturbation can source $\zeta$ if the equation of state depends on the charge fraction.

At linear order, write a transfer matrix between an initial time after diffusive freeze-out and a final radiation epoch,
\begin{equation}
 \begin{pmatrix}\zeta_f\\S_f\end{pmatrix}
 =
 \begin{pmatrix}1&T_{\zeta S}(k)\\0&T_{SS}(k)\end{pmatrix}
 \begin{pmatrix}\zeta_i\\S_i\end{pmatrix}.
 \label{eq:transfer}
\end{equation}
The final spectra are
\begin{align}
 \Pz^{f}&=\Pz^{i}+2T_{\zeta S}\mathcal P_{\zeta S}^{i}
 +T_{\zeta S}^2\PS^{i},\\
 \mathcal P_{\zeta S}^{f}&=T_{SS}
 \left(\mathcal P_{\zeta S}^{i}+T_{\zeta S}\PS^{i}\right),\\
 \PS^{f}&=T_{SS}^2\PS^{i}.
 \label{eq:transferredspectra}
\end{align}
A chemical transition, decay of a charge-carrying species, or dark-sector freeze-out can generate $T_{\zeta S}\neq0$.

The terms linear in $T_{\zeta S}$ describe interference between initially correlated curvature and entropy modes. They can raise or lower the final curvature power depending on the sign of $\mathcal P_{\zeta S}^i$, so a correlated entropy mode is not equivalent to adding an independent positive spectrum. Isocurvature may also be generated from initially adiabatic fluctuations when species depart from equilibrium; a separate-universe treatment of thermal dark-matter freeze-in and freeze-out shows that this effect is generally suppressed on super-Hubble scales but can be calculated systematically \cite{Holst:2024dmiso}.

For the thermal-seeding mechanism considered here, the initial cross-spectrum need not vanish because $\Sigma_{\rho n}\neq0$. This is a qualitative distinction from phenomenological analyses that assume statistically independent adiabatic and isocurvature modes.

\subsection{Transfer coefficient from a slowly varying charged component}

Combining Eqs.~\eqref{eq:zetadot} and \eqref{eq:pnadS}, and neglecting gradients, gives
\begin{equation}
 \frac{\dd\zeta}{\dd N}=-\frac{n}{\rho+p}
 \left(\frac{\partial p}{\partial n}\right)_\rho S,
 \label{eq:zetadN}
\end{equation}
where $N=\ln a$ is the number of e-folds.  If $S$ is approximately conserved, the transfer function is estimated analytically as
\begin{equation}
 T_{\zeta S}(N_f,N_i)
 \simeq-\int_{N_i}^{N_f}\dd N\,
 \frac{n}{\rho+p}
 \left(\frac{\partial p}{\partial n}\right)_\rho.
 \label{eq:TzSestimate}
\end{equation}
Conversion is therefore localized at epochs where the charged component affects the pressure: a mass threshold, decay, annihilation, or phase transition.  In an exactly conformal epoch the integrand vanishes.  If conversion occurs over $\Delta N$ e-folds with nearly constant coefficient $\gamma_S$, then $T_{\zeta S}\simeq-\gamma_S\Delta N$.  Order-one conversion requires either a dynamically important charged component or a prolonged conversion interval.

The coefficient inside the integral is dimensionless. It vanishes in the conformal limit and becomes appreciable only when composition affects pressure at fixed energy. Equation~\eqref{eq:TzSestimate} therefore identifies the epochs that must be resolved in a numerical multifluid calculation.

\subsection{Decay estimate}

As a simple limiting case, let a nonrelativistic charged species $X$ carry curvature $\zeta_X$ and coexist with radiation of curvature $\zeta_r$.  Immediately before a sudden decay, define
\begin{equation}
 r_D\equiv\frac{3\rho_X}{4\rho_r+3\rho_X}.
\end{equation}
Energy conservation on the decay hypersurface gives, at linear order,
\begin{equation}
 \zeta_f\simeq(1-r_D)\zeta_r+r_D\zeta_X,
 \qquad
 \Delta\zeta\simeq\frac{r_D}{3}S_{Xr},
\end{equation}
where $S_{Xr}=3(\zeta_X-\zeta_r)$.  Thus $T_{\zeta S}\simeq r_D/3$.

The parameter $r_D$ is an enthalpy-weighted energy fraction evaluated immediately before decay. It approaches zero for a negligible decaying component and unity when the nonrelativistic species dominates. The factor $1/3$ follows from the convention $S_{Xr}=3(\zeta_X-\zeta_r)$.  The same charge fluctuation is weakly imprinted when $X$ remains subdominant, but can be efficiently converted if $X$ temporarily carries an appreciable fraction of the total energy.  This weighting is the linear limit of the standard sudden-decay matching calculation used in the curvaton literature; fully nonlinear and finite-duration corrections were quantified in Ref.~\cite{Sasaki:2006decay}.

\label{sec:obs}

\subsection{Correlated isocurvature on CMB scales}

A convenient phenomenological parametrization at a pivot scale $k_\star$ is

The fraction $\beta_{\rm iso}$ lies between zero and one when the auto-spectra are positive. The correlation coefficient $\cos\Delta$ lies between minus one and one by covariance positivity; its sign fixes whether curvature and entropy perturbations interfere constructively or destructively after transfer.
\begin{equation}
 \beta_{\rm iso}(k_\star)=
 \frac{\PS(k_\star)}{\Pz(k_\star)+\PS(k_\star)},
 \qquad
 \cos\Delta(k_\star)=
 \frac{\mathcal P_{\zeta S}}
 {\sqrt{\Pz\PS}}.
 \label{eq:phenoparams}
\end{equation}
The framework predicts both quantities from the same susceptibility matrix and transfer functions. Since large-scale data strongly constrain nonadiabatic initial conditions, a viable CMB scale realization requires either a small thermal charge fraction, efficient damping $T_{SS}\ll1$, or a blue spectrum that suppresses power at the CMB pivot. Gauge-invariant multifluid perturbation theory and standard definitions of adiabatic and entropy modes are reviewed in Ref.~\cite{Malik:2008im}.

\section{Discussion}
\label{sec:discussion}

The combined framework separates three ingredients that are often conflated in thermal-seeding scenarios.

The logic of the calculation may be summarized as follows. First, an equation of state fixes $\Sigma$ and the higher cumulants. Second, transport and the background determine the mode-dependent freeze-out time. Third, projection and transfer convert the frozen thermodynamic variables into $\zeta$ and $S$. Finally, a Boltzmann evolution would map these primordial spectra into observable temperature, polarization, and matter correlations. Equilibrium thermodynamics fixes the equal-time covariance and the hierarchy of connected cumulants. Transport determines how fluctuations approach equilibrium and when they cease to track it. Gravitational evolution maps the frozen thermal variables into curvature and entropy perturbations. This separation makes the assumptions of each result transparent and allows a microscopic model to be tested stage by stage.
At finite chemical potential, the same pressure function determines energy, charge and mixed cumulants, providing a common thermodynamic basis for studying scalar power, isocurvature and higher-point correlations.
This makes the higher-point functions sensitive to the thermal history: near a phase transition, changes in the higher derivatives of the pressure can enhance non-Gaussianity, as is the case in cyclic inflation \cite{Biswas:2013phase}.

For an isolated conformal sector with conserved charge-to-entropy ratio, the susceptibility matrix scales as a fixed power of temperature. Diffusive freeze-out then produces the universal scaling $\mathcal P_S\propto k^3$. Changing only the power-law temperature dependence of the diffusion coefficient cannot remove this blue behavior. A different spectrum requires nonconformal thermodynamics, a transition or threshold, a nonstandard background, nonextensive correlations, or a departure from the conserved trajectory. A $k^3$ curvature spectrum also arises when radiation-temperature fluctuations determine the end of thermal inflation \cite{LythStewart:1996thermal,Bae:2025thermal}.
In this example, energy injection compensates almost entirely for the dilution of the bath, so $T$ evolves slowly along with $H$ and $\mathcal A_X$. The scaling $\mathcal P_{\zeta_X}\propto\mathcal A_X^2TH$ then explains the weak scale dependence. Such a slowly evolving bath is maintained by dissipative interactions in warm inflation \cite{Berera:1995warm,Hall:2004warm}, suggesting a natural setting for a microscopic realization of the assumed energy source.

The open dark-fermion example of Sec.~\ref{sec:dark_fermion} follows a sourced quasi-de Sitter trajectory with constant physical $\mu$ and constant $\Omega_X$. For the representative choice $\mu/T_\star=2$ and $\Omega_X=0.003$, fixing the background evolution and thermal scale reproduces $\mathcal P_\zeta(k_\star)=2.10\times10^{-9}$ and $n_s(k_\star)=0.9649$. The agreement extends beyond the pivot: over $0.002\leq k/k_\star\leq4$, the spectrum differs from the corresponding power law by at most approximately $0.30\%$. Thus, within the adopted matching prescription, the example yields a nearly scale-invariant, red scalar spectrum over an extended interval.  Figures~\ref{fig:pzeta} and \ref{fig:spectral_diagnostics} establish that this agreement is accompanied by a smooth residual and weak running. The two-helicity tensor ratio and Fig.~\ref{fig:ng} show that the tensor contribution and nonlinear cumulant amplitudes remain small. Collectively, the figures indicate that the benchmark behavior is controlled by a slowly evolving sourced background rather than by a narrow spectral feature.

Once these scalar targets and the representative inputs are fixed, the same construction gives $f_{\rm NL}^{}(k_\star)\simeq0.1055$ and $g_{\rm NL}^{}(k_\star)\simeq0.01583$. Assigning the independent vacuum tensor spectrum also gives $r_{t/s}(k_\star)\simeq1.82\times10^{-4}$. The quoted higher-order amplitudes retain the normalization and matching assumptions used in the example; comparison with observational non-Gaussianity templates additionally requires the momentum dependence and subsequent perturbation transfer.
Finite $\mu$ modifies the thermodynamic cumulants and their evolution along this trajectory. Maintaining the representative solution requires both energy and charge exchange with a reservoir, whose perturbations must ultimately be included in the evolution of the total curvature perturbation.

The hydrodynamic approximation introduces a separate set of consistency requirements. The microscopic equilibration time and current-relaxation time must remain shorter than the Hubble time. In the diffusive channel, the correlation length must be smaller than the diffusion length, which must itself remain sub-Hubble. If these hierarchies fail, a kinetic or causal stochastic treatment is required. These restrictions are collected in Appendix~\ref{app:validity}.
\section{Conclusions}
\label{sec:conclusion}

The thermal mechanism studied here determines the origin and statistics of primordial perturbations, but it does not by itself explain why the observable universe is so large, homogeneous, and nearly spatially flat. A complete scenario must embed these fluctuations in an appropriate early-universe background, such as a sufficiently long inflationary phase or a contracting phase followed by a nonsingular bounce \cite{Guth:1981inflation,Lemoine:2008zz,BrandenbergerPeter:2017bounce}.

The main result of our analysis is a direct link between equilibrium statistical mechanics and primordial cosmological correlators. At finite chemical potential, thermal fluctuations are intrinsically multivariate: energy and charge fluctuate together, and their full grand-canonical susceptibility matrix determines the curvature, charge isocurvature, and cross-correlation amplitudes. This formulation identifies the equation of state, transport coefficients, and cosmological transfer history as distinct physical inputs, thereby turning thermal seeding into a sequence of calculations that can be tested independently.

The first result is a general constraint. For an isolated, adiabatic, extensive conformal sector with conserved charge-to-entropy ratio and power-law diffusion, diffusive freeze-out gives
\begin{equation}
 \mathcal P_S(k)\propto k^3,
 \qquad n_{\rm iso}=4,
\end{equation}
independently of the temperature exponent of the diffusion coefficient. In the same limit, the equal-time local-equilibrium curvature-isocurvature covariance vanishes exactly in the conformal source basis; later transport or conversion can nevertheless generate a final cross-spectrum. These results show that neither finite chemical potential nor a change from Hubble crossing to diffusion crossing is sufficient by itself to produce a nearly scale-invariant spectrum. The universal blue scaling is therefore not merely a feature of one example: it is a sharp guide to which assumptions must be relaxed in any successful thermal construction.

The second result is constructive. Once the thermal sector is treated as an open subsystem during a sourced quasi-de Sitter phase, the conformal no-go conditions no longer apply. For the representative massless dark-fermion solution with $\mu/T_\star=2$ and $\Omega_X=0.003$, the scalar amplitude and tilt are reproduced at the pivot,
\begin{equation}
 \mathcal P_{\zeta_X}(k_\star)=2.10\times10^{-9},
 \qquad n_s(k_\star)=0.9649,
\end{equation}
while the spectrum remains within approximately $0.30\%$ of the corresponding power law over $0.002\leq k/k_\star\leq4$. The same thermodynamic cumulant hierarchy yields a small negative running, $\alpha_s(k_\star)=-1.67\times10^{-4}$, together with small positive local cumulant amplitudes, $f_{\rm NL}^{}(k_\star)\simeq0.1055$ and $g_{\rm NL}^{}(k_\star)\simeq0.01583$. If an independent vacuum tensor spectrum is imposed, the tensor-to-scalar ratio is $r_{t/s}(k_\star)\simeq1.82\times10^{-4}$. The importance of this result is that the scalar spectrum, its running, and its higher-order correlations all descend from the same finite-temperature equation of state and the same background trajectory rather than from unrelated phenomenological inputs.

The numerical solution also shows what the mechanism requires dynamically. Keeping $\Omega_X$ and the physical chemical potential nearly constant requires continuous energy and charge transfer, with $Q_E/(H\rho_X)\simeq3.96$ and $Q_N/(Hn_X)\simeq2.98$ at the pivot. Thus the successful red spectrum is driven primarily by the sourced, slowly evolving background, while finite $\mu$ enriches the thermodynamic structure and modifies the scale dependence. 
These results suggest a concrete future outlook to derive the source terms and local equilibration rates from microscopic interactions, include reservoir fluctuations in the coupled gauge-invariant perturbation equations, and propagate the correlated curvature and isocurvature modes through the subsequent cosmological history. It will determine the final momentum dependence of the bispectrum and trispectrum and enable direct confrontation with temperature, polarization, spectral-distortion, and small-scale-structure observables. The principal conclusion is therefore the isolated conformal route is decisively constrained, but a sourced thermal sector can generate realistic, weakly running primordial scalar correlations with calculable higher-order structure. This opens a well-defined alternative route by which early-universe statistical fluctuations can become observable cosmological initial conditions.

\section*{Acknowledgement}
A.~G. acknowledges support
from the Royal Society, UK, Fellowship funding reference: NIF\ R1\ 253963. A.~M. is supported by the Natural Sciences and Engineering Research Council of Canada (NSERC).

\appendix
\section{Numerical scales and present-day frequency}
\label{app:scales}

Here $a_k$ is the scale factor at diffusive freeze-out, $a_0$ is its present value, $T_0$ is the present photon temperature, and $g_{\ast s}$ counts effective entropy degrees of freedom. The estimate assumes no entropy production after freeze-out; any later entropy release rescales the frequency through the ratio of $g_{\ast s}$.

A comoving mode freezing at temperature $T_k$ has
\begin{equation}
 \frac{k}{a_0}=\frac{a_k}{a_0}\sqrt{\frac{c_DH_k}{D_k}},
 \qquad
 \frac{a_k}{a_0}=\frac{T_0}{T_k}
 \left(\frac{g_{*s,0}}{g_{*s,k}}\right)^{1/3},
 \label{eq:redshiftmap}
\end{equation}
assuming entropy conservation after freeze-out.  The corresponding present frequency is $f_0=k/(2\pi a_0)$.  For radiation domination and $D=d_D/T$,
\begin{equation}
 f_0=\frac{T_0}{2\pi}
 \left(\frac{g_{*s,0}}{g_{*s,k}}\right)^{1/3}
 \left(\frac{c_D\sqrt{\pi^2g_*/90}}{d_D}\frac{T_k}{M_{\rm Pl}}\right)^{1/2}.
 \label{eq:frequencyestimate}
\end{equation}
Unlike horizon-crossing signals, the frequency scales as $T_k^{1/2}$ for conformal diffusion rather than linearly with $T_k$.  This modified map is important when connecting a thermal feature to spectral distortions, small-scale structure, pulsar timing, or interferometers.

\section{Regime-of-validity checklist}
\label{app:validity}

The microscopic equilibration time is denoted by $\tau_{\rm mic}$, the current-relaxation time by $\tau_J$, the correlation length by $\xi$, and the diffusion length by $\ell_D$. The diffusion eigenvalue $D_r$ must be positive. Each inequality tests a distinct approximation, so satisfying only the final linearity condition is insufficient.

A microscopic realization must satisfy the following hierarchy of scales:
\begin{align}
 &H\tau_{\rm mic}\ll1 &&\text{local thermal equilibrium},\\
 &H\tau_J\ll1 &&\text{first-order diffusion limit},\\
 &k_{\rm ph}\xi\ll1 &&\text{hydrodynamic gradient expansion},\\
 &\xi\ll\ell_D\ll H^{-1} &&\text{many cells and sub-Hubble freeze-out},\\
 &\mathcal P_S\ll1 &&\text{linear perturbation theory},\\
 &\Sigma\succeq0,\quad D_r>0 &&\text{thermodynamic and transport stability}.
\end{align}
Failure of the first three conditions does not necessarily eliminate the model, but it invalidates the equilibrium Markovian formulas and requires a kinetic or causal stochastic calculation.

\section{Multiple conserved charges}
\label{app:multi}

Indices $a,b=1,\ldots,N$ label conserved charges, whereas $r$ labels eigenmodes of the diffusion operator. The matrix $R$ rotates from the original charge basis into the transport eigenbasis. Because this rotation need not diagonalize the susceptibility matrix, statistically correlated eigenmodes can freeze at different temperatures.

For $N$ conserved charges, let
\begin{equation}
 \delta\bm q=(\delta\rho,\delta n_1,\ldots,\delta n_N)^T.
\end{equation}
The susceptibility matrix is $(N+1)\times(N+1)$. Define
\begin{equation}
 S_a=\bm u_a^T\delta\bm q,
 \qquad
 (\bm u_a)^T=
 \left(-\frac{1}{\rho+p},0,\ldots,\frac{1}{n_a},\ldots,0\right).
\end{equation}
Then
\begin{equation}
 \mathcal C_{S_aS_b}=\bm u_a^T\Sigma\bm u_b.
\end{equation}
Transport is controlled by a diffusion matrix $D_{ab}$. Its eigenvectors, not necessarily the original charge basis, are the modes that freeze independently. If $R$ diagonalizes the linearized diffusion operator, the freeze-out condition for eigenmode $r$ is
\begin{equation}
 D_r(T)\frac{k^2}{a^2}=c_rH.
\end{equation}
Because the thermodynamic and diffusion matrices need not commute, the isocurvature correlation angle can be scale dependent even when all equilibrium susceptibilities are smooth.  Complete baryon-electric-strangeness transport calculations provide explicit examples in which off-diagonal diffusion entries are phenomenologically important \cite{Greif:2018diffusion}.

\section{Window functions and spectral normalization}
\label{app:window}

The window $W_R$ is normalized to unity and has width $R$ in physical coordinates. The effective volume is defined by $(V_R^{\rm eff})^{-1}=\int\dd^3x\,W_R^2$, which is the volume entering the variance of a smoothed white-noise field.

For a Gaussian physical-space window
\begin{equation}
 W_R(\bm x)=\frac{1}{(2\pi R^2)^{3/2}}
 \exp\left(-\frac{x^2}{2R^2}\right),
\end{equation}
we find
\begin{equation}
 \int\dd^3x\,W_R^2=\frac{1}{8\pi^{3/2}R^3},
 \qquad
 V_R^{\rm eff}=8\pi^{3/2}R^3.
\end{equation}
Order-one factors in the sudden-freeze-out amplitude depend on this choice and on the precise matching criterion $c_D$. The tilt result in Eq.~\eqref{eq:niso4} is independent of these constants.

\section{Weakly coupled relativistic gas at small chemical potential}
\label{app:gas}

The coefficients $c_0$, $c_2$, and $c_4$ are dimensionless equation-of-state coefficients. Charge-conjugation symmetry makes the pressure even in $x=\mu/T$, so only even powers appear. The susceptibility is positive when the leading coefficient $c_2$ is positive.

For a relativistic species at small $x=\mu/T$, write
\begin{equation}
 p=T^4\left(c_0+c_2x^2+c_4x^4+\cdots\right).
\end{equation}
Then
\begin{equation}
 n=T^3\left(2c_2x+4c_4x^3+\cdots\right),
 \qquad
 \rho=3p.
\end{equation}
The charge susceptibility at fixed temperature is
\begin{equation}
 \chi=\left(\frac{\partial n}{\partial\mu}\right)_T
 =T^2\left(2c_2+12c_4x^2+\cdots\right).
\end{equation}
At exactly vanishing background charge, $n=0$, the fractional variable $\delta n/n$ is singular. The physically appropriate isocurvature variable is then a charge yield perturbation normalized to entropy, $\delta(n/s)$, or the energy density of the eventual charge-carrying relic. The formalism in the main text assumes a nonzero homogeneous $n$; the zero-asymmetry case must be treated with this alternative normalization.

\end{document}